\documentclass[lettersize,journal]{IEEEtran}

\usepackage{amsmath,amsfonts}
\usepackage{algorithmic}
\usepackage{array}
\usepackage{hyperref}
\usepackage{textcomp}
\usepackage{stfloats}
\usepackage{url}
\usepackage{verbatim}
\usepackage{graphicx}
\usepackage{cite}
\usepackage {threeparttable}
\usepackage{longtable}
\usepackage{bbding}
\usepackage{pifont}
\usepackage{booktabs}
\usepackage{makecell}
\usepackage[ruled,vlined]{algorithm2e}
\usepackage{enumitem}
\usepackage{amssymb}
\usepackage{booktabs}
\usepackage{subcaption}
\usepackage{multirow} 
\usepackage{color}

\newcommand{\mc}[1]{\mathcal{#1}}
\newcommand{\dt}[1]{{#1}^{\prime}}

\begin{document}

\title{Electricity Cost Minimization for Multi-Workflow Allocation in Geo-Distributed Data Centers}

\author{Shuang Wang, He Zhang, Tianxing Wu*, Yueyou Zhang, Wei Emma Zhang, and Quan Z. Sheng
\thanks{Manuscript received ; revised }
\thanks{Corresponding author: Tianxing Wu (tianxingwu@seu.edu.cn) }
\thanks{This work is supported by the National Key Research and Development Program (No. 2022YFF0902800), Natural Science Foundation of Jiangsu Province (No. BK20220803), NSFC (No. 62302095, 62376058), Southeast University Interdisciplinary Research Program for Young Scholars, 
 and the Big Data Computing Center of Southeast University.
}
 \thanks{S. Wang, H. Zhang, T. Wu, and Y. Zhang are with the School of Computer Science and Engineering, Southeast University, Nanjing, 211189, China.
%
E-mail:\{shuangwang, tianxiangwu\}@seu.edu.cn, 
zhanghezhedu@163.com, is\_zhangyy@163.com.
\\
\IEEEcompsocthanksitem W. E. Zhang is 
with the School
of Computer and Mathematical Sciences, and 
the Australian Institute for Machine
Learning, The University of Adelaide, Australia. Email: wei.e.zhang@adelaide.edu.au.\\
 Q. Z. Sheng is with the School of Computing, Macquarie University, Sydney, NSW 2109,  Australia.
E-mail: michael.sheng@mq.edu.au.}
}

\markboth{IEEE Transactions on Services Computing,~Vol.~xx, No.~x, November~2024}%
{Shell \MakeLowercase{\textit{et al.}}: A Sample Article Using IEEEtran.cls for IEEE Journals}


\maketitle

\begin{abstract}
Worldwide, Geo-distributed Data Centers (GDCs) provide computing and storage services for massive workflow applications, resulting in high electricity costs that vary depending on geographical locations and time. How to reduce electricity costs while satisfying the deadline constraints of workflow applications is important in GDCs, which is determined by the execution time of servers,  power, and  electricity  price. Determining the completion time of workflows with different server frequencies can be challenging, especially in scenarios with heterogeneous computing resources in GDCs. Moreover, the electricity price is also different in geographical locations and may change dynamically. To address these challenges, we develop a geo-distributed system architecture and propose an Electricity Cost aware Multiple Workflows Scheduling algorithm (ECMWS) for servers of GDCs with fixed  frequency and power. ECMWS comprises four stages, namely workflow sequencing, deadline partitioning, task sequencing, and resource allocation where two graph embedding models and a policy network are constructed to solve the Markov Decision Process (MDP). After statistically calibrating parameters and algorithm components over a comprehensive set of workflow instances, the proposed algorithms are compared with the state-of-the-art methods over two types of workflow instances. The experimental results demonstrate that our proposed algorithm significantly outperforms other algorithms, achieving an improvement of over 15\% while maintaining an acceptable computational time. The source codes are available at \hyperlink{https://gitee.com/public-artifacts/ecmws-experiments}{https://gitee.com/public-artifacts/ecmws-experiments}.
\end{abstract}

\vspace{-2mm}
\section{Introduction}
\IEEEPARstart{W}{ith} the advancement of global digital transformation, GDCs have gained widespread adoption.
These data centers are located in different geographical locations, access and transfer data to each other through the network, and provide efficient and reliable services for users around the world. GDCs have advantages in  high fault tolerance and availability, 
as well as the ability to tailor services based on user requirements \cite{7373667,7346506,6587037}. Many applications deployed in cross-domain data centers are represented as directed acyclic graphs (DAGs), which are submitted to systems named workflows. For instance, in  scientific computing, Montage \cite{10.1504/IJCSE.2009.026999} is an astronomical image mosaic application that involves a workflow with multiple image processing and data analysis tasks. 
Empirical data indicate that the global energy expenditure attributed to data centers ranges between 2\% and 5\%, exhibiting a secular upward trajectory \cite{IEA2022}. However, a significant amount of electricity is consumed in cross-domain data centers when executing these applications \cite{10417832,8954766}. Although achievements have been made to improve efficiency \cite{9662214,10001941}, how to reduce  electricity cost in GDCs is urgent and necessary.
In this paper, we consider a multi-workflow scheduling problem for minimizing electricity costs in a cross-domain data center environment.



Electricity pricing varies 
across geographically dispersed server locations. Concurrently, the processing of tasks on such distributed servers necessitates the consideration of transmission latency, thereby augmenting the complexity of the optimization problem. For workflow scheduling, the orchestration of computational tasks within distributed data centers is complex, particularly when dealing with the constraints of deadlines and the imperative to optimize energy consumption. The challenges associated with resource allocation in such scenarios are multifaceted 
with 
the following issues:
\begin{itemize}
\item The sharing of resources across data centers for multiple workflows makes it difficult to predict completion time and design algorithms that meet deadline constraints. Resource limitations and heterogeneity among data centers lead to imbalanced task execution on varied servers, complicating the scheduling process.
\item The optimization of electricity costs is hindered by the trade-off between the economic use of electricity and the availability of computational resources. Scheduling tasks to cheaper electricity centers may increase execution time, while scheduling to resource-rich centers may reduce execution time but increase electricity costs.

\item The competition for resources  among workflows intensifies the conflict between resource abundance and electricity cost-effectiveness. Aggressive cost reduction strategies may lead to suboptimal resource allocation, potentially increasing total electricity cost due to the allocation of tasks to more expensive or slower data centers.
\end{itemize}





Tackling the aforementioned 
challenges, the main contributions made in this work are summarized as follows:
\begin{itemize}
    \item We introduce a novel architecture for scheduling multiple workflows across data centers. It incorporates the heterogeneity of computational resources and the dynamic nature of electricity pricing, leading to the development of foundational mathematical models.
    \item We propose an Electricity Cost aware Multiple Workflows Scheduling algorithm (ECMWS) to minimize electricity costs under fixed server frequency and power constraints which include four-stage process: workflow sequencing, sub-deadline partitioning, task sequencing, and resource allocation using graph embedding. 
    \item To allocate resources efficiently, we use the Actor-critic framework for workflows by embedding graphs according to which suitable DCs and servers are determined by Deadline Assured Resource Allocation algorithms. The experiments validate that the proposed algorithm outperforms the other algorithms  over  15\% on the electricity cost  in  GDCs.
  
\end{itemize}


 The 
 remainder of this paper is organized as follows.
 The related research is reviewed and compared in Section \ref{sec2}.
 The mathematical model is defined for the  workflow scheduling problem in GDCs in Section \ref{sec3}. In Section \ref{sec4}, we 
 present the 
 ECMWS algorithm to minimize the electricity  cost. The experimental results with real-world datasets are analyzed and reported in Section \ref{sec5}. Finally, the conclusion is 
 summarized in Section \ref{sec6} with several highlighting remarks.

\vspace{-2mm} 
\section{Related Work}\label{sec2}
\vspace{-1mm}
Scheduling workflows in  cross-domain DCs is  difficult, as it has been proven to be NP-hard in the previous study \cite{9662214}. Energy-saving task scheduling in data center scenarios has been widely studied. Wakar et al. \cite{Ahmad2021} pioneered the study of workflow scheduling under budget constraints, aiming to minimize energy consumption.
Furthermore, Li et al. \cite{8954766} under the assumption of fixed bandwidth constraints between DCs, balanced energy cost with electricity prices  for scheduling workflows.
However, most of these studies \cite{9662214,Ahmad2021,8954766} ignore the fluctuating electricity prices associated with both location and time for multiple workflows across various geographical locations. In \cite{10.1007/978-981-97-5760-2_3}, the  Dynamic Voltage and Frequency Scaling technology was adopted to save electricity cost by training a policy network for resource allocation while in this paper, we focus on the dynamic price of electricity for geo-distributed data centers with server frequency
and power to improve resource allocation.

Compared to the single-workflow scheduling, several studies \cite{10.1007/s10586-020-03079-1,9705055,10.1007/s00607-021-00930-0,10382635} 
focused on multi-workflow scheduling that addressed more complex optimization goals, such as total completion time, cost, energy use, and QoS (Quality of Service) for multiple workflows in cloud computing. For instance, Rizvi et al. \cite{10.1007/s10586-020-03079-1} proposed a fair budget-constrained multi-workflow scheduling algorithm which balanced time and cost by adjusting cost-time efficiency factors. Multi-workflow scheduling is also amenable to application across a variety of computational scenarios. For task scheduling in  IoT (Internet of Things), Attiya et al. \cite{9705055} enhanced the local search by the Salp Swarm Algorithm, which effectively reduced the convergence time of the global optimal solution. 
In fog-cloud environments, Samia et al. \cite{10.1007/s00607-021-00930-0} proposed a two-stage scheduling algorithm that minimized workflow energy consumption and makespan. In multi-tenant WaaS (Workflow as a Service) environments, Rodriguez et al. \cite{10.1016/j.future.2017.05.009} proposed a heuristic dynamic scheduling algorithm  catering to both short-term and long-term resource demands of different workflows. These studies provide valuable insights for workflow scheduling in cloud environments. However, the spatio-temporal diversity of electricity prices and the geo-distributed characteristics are ignored. 


Deep Reinforcement Learning (DRL) has emerged as a pivotal approach in workflow scheduling \cite{ZHANG2023120972,YAN2022107688,9592509}. Unlike conventional techniques, DRL is capable of making real-time decisions, which significantly reduces the computational cost. 
Zhang et al. \cite{ZHANG2023120972} proposed a real-time workflow scheduling method that combines Genetic Algorithm (GA) with DRL to reduce 
execution cost and response time. Furthermore, there is a growing interest in  DRL-based scheduling algorithms to minimize energy \cite{YAN2022107688,9592509}. 
Yan et al. \cite{YAN2022107688} proposed a DRL based job scheduling algorithm to reduce energy consumption for job execution 
while keeping high QoS. Additionally, the Actor-Critic framework, a key component in many DRL algorithms, has been widely adopted for dynamic workflow scheduling \cite{9592509}. 
Although these methods have shown some promise, they  have not fully considered the geographical distributed characteristics on data center locations.

 Energy-efficient  workflow scheduling problems are related to electricity cost. However, due to the dynamic prices in geo-distributed DCs, scheduling multiple workflows to minimize total electricity cost is more complex than  energy minimization problems because of the dynamic electricity characteristic in GDCs. 
 In this paper, we aim to minimize the electricity cost for effectively scheduling multiple workflows in a distributed DC environment for the diverse electricity prices across different locations and time periods.

\section{System Framework and Problem Descriptions}\label{sec3}
For the considered problem, $N$ workflow applications with deadline constraints randomly arrive at the system, i.e., ${G}=\left\{G_1,\dots,G_N\right\}$. Each workflow $G_w\in {G}$ is represented as a DAG, denoted as $G_w=(V_w, E_w)$ where $V_w=\left\{v^w_1,\dots,v^w_{n_w}\right\}$ is the set of tasks, $n_w$ is the total number of tasks in workflow $G_w$, and $v^w_i\in V_w$ is the $i^{th}$ task of workflow $G_w$. Each task $v^w_i$ has a workload $W_{w,i}$, representing the number of instructions to be executed. Each task $v^w_i$ has direct predecessor and successor task sets $PR^w_i$ and $SU^w_i$. Note that the ${PR}_i^w$ set is empty if the task is the first in the workflow. Each task $v^w_i$ needs to receive data produced by its predecessor tasks $v^w_{i'}\in PR^w_i$ before execution, with the data size $S^w_{i', i}$. Table \ref{tbl:notation} presents  important notations   in this paper.

\vspace{-2mm} 
\subsection{System Architecture}
\begin{figure}[!bt]
    \centering
\includegraphics[width=0.45\textwidth]{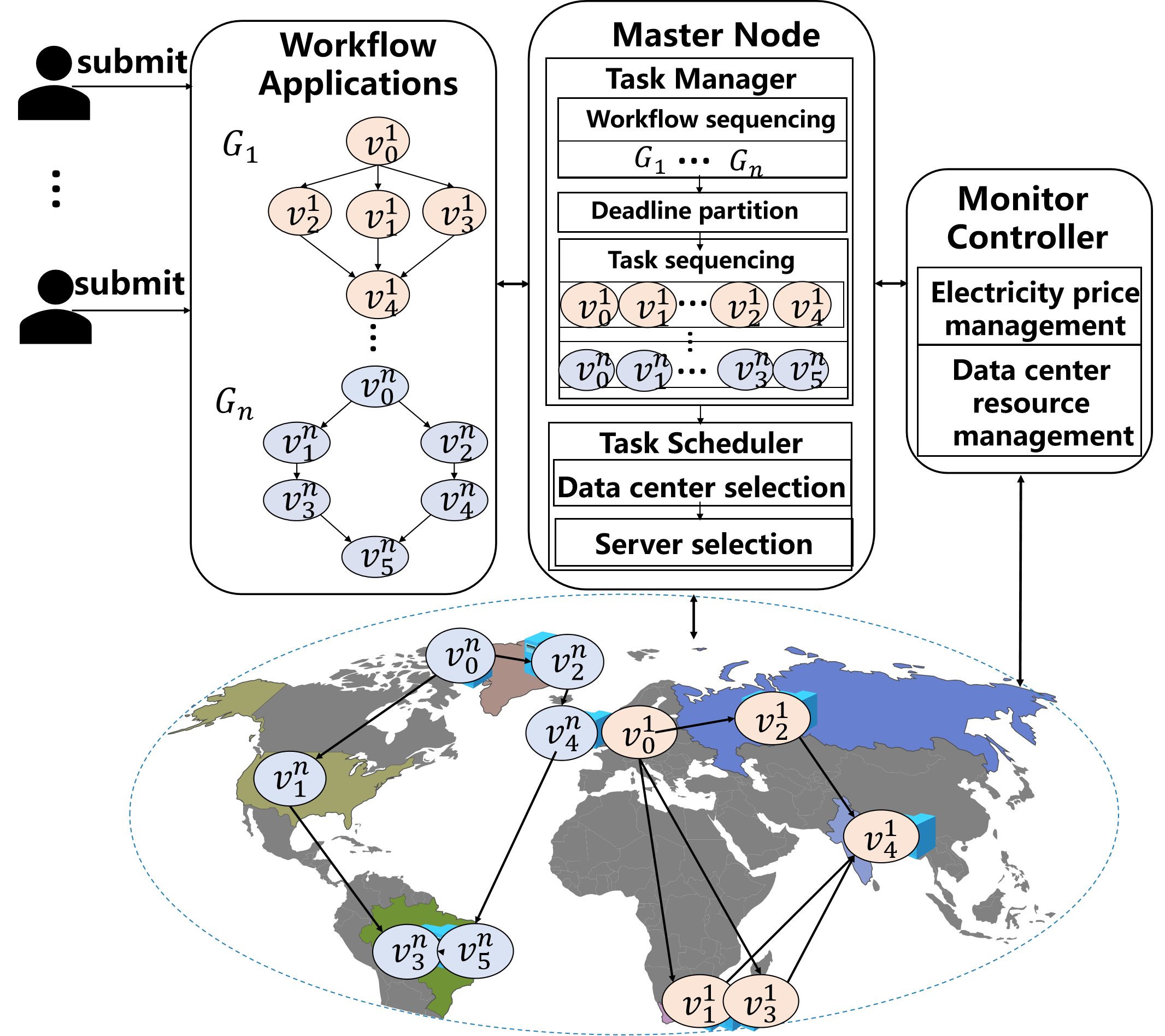}
    \caption{System Architecture.}
    \vspace{-5mm}
    \label{fig:sys_arch}
\end{figure}

\vspace{-1mm} 
To minimize the electricity cost, the system architecture of multiple workflows scheduling in geo-distributed DCs is shown in Fig. \ref{fig:sys_arch}.
The master node receives multiple workflows submitted by users at each time slot, and at the end of each slot, the master node schedules all workflows in the previous slot. The master node has two modules: {\em Task Manager} and {\em Task Scheduler}. 
The Task Manager sorts a batch of workflows to be scheduled in each slot (Workflow Sequencing),  divides the sub-deadlines (Deadline Partition) for each task in the workflow sequentially and determines the scheduling sequence (Task Sequencing). The Task Scheduler sequentially selects a DC and a server 
for each task in the scheduling sequence. The Monitor Controller is responsible for managing resources in the geo-distributed DCs (DC Resource Management) and the electricity prices in corresponding locations (Electricity Prices Management). After the tasks are executed in the geo-distributed DCs, the performance results 
are returned to the users through the master node.

We assume that the service provider operates $M$ DCs: ${D}=\left\{{D}_1,\dots,{D}_M\right\}$, where a DC ${D}_k\in {D}$ has an electricity price $p_k(t)$ that periodically changes with time $t$. The $k^{th}$ DC $\mc{D}_k$  contains $\mu_k$ clusters, where the $j^{th}$ cluster
is represented as ${H}^k_j$. 
Cluster $\mc{H}^k_j$ has $\omega^k_j$ 
servers. Let $\omega^k$ denote the total number of 
servers in each DC:
 $\omega^k=\sum_{j=1}^{\mu_k}\omega^k_j$.
$\overline{\omega}$ is denoted as the total number of 
servers in the system:
 $\overline{\omega}=\sum_{k=1}^M\omega^k$.
The processing frequency and power of a 
server $\mc{V}^k_{j,l}$ are represented as $\xi_{k,j,l}$ and $P_{k,j,l}$, respectively. The runtime of the system is computed by a series of continuous slots. During each time slot, the master node accepts the workflows and schedules all these workflows at the end of each time slot. 
	
	\begin{table}[!h]
		\caption{Important notation definition}
		\label{tbl:notation}
		\centering
		\begin{tabular}{cl}
			\hline
			Symbol & Definition   \\
			\hline
			$N$ & Total number of workflow applications \\
			$M$ & Total number of data centers \\ 
			$G_w$ & The $w^{th}$ workflow application \\ 
			$T^{submit}_w$ & Submission time of application $G_w$ \\
			$v^w_i$ & The $i^{th}$ task in application $G_w$ \\
			$n_w$ & Number of tasks in the workflow $G_w$ \\
			$\mc{D}_k$ & The $k^{th}$ data center \\
			$d_w$ & Deadline of application $G_w$ \\
			$\mc{H}^k_j$ & The $j^{th}$ cluster
            in data center $\mc{D}_k$ \\
			$\mc{V}^k_{j,l}$ & The $l^{th}$ server
        in cluster 
            $\mc{H}^k_j$ \\
			$\mu_k$ & Number of clusters 
            in data center $\mc{D}_k$ \\ 
            $\omega^k_j$ & Number of servers in cluster $\mc{H}^k_j$ \\ 
			$\overline{\omega}$ & Number of servers
            in all data centers \\ 
			$\xi_{k,j,l}$ & Processing speed (MIPS) of servers
            $\mc{V}^k_{j,l}$ \\ 
			$P_{k,j,l}$ & Power of servers
            $\mc{V}^k_{j,l}$ \\ 
			$W_{w,i}$ & Workload (MI) of task $v^w_i$ \\
			$PR^w_i$ & Set of predecessor tasks of task $v^w_i$ \\
			$SU^w_i$ & Set of successor tasks of task $v^w_i$ \\
			$S^w_{i',i}$ & Data volume to be transferred from $v^w_{i^{'}}$ to $v^w_i$ \\
			$T^{work}_{w,i}$ & Processing time of task $v^w_i$ \\ 
			$T^{trans}_{w,\dt{i},i}$ & Data transmission time from $v^w_{\dt{i}}$ to $v^w_i$ \\
			$p_k(t)$ & Electricity price of data center $\mc{D}_k$ at time $t$ \\
			$C_{w,i}$ & Electricity cost of task $v^w_i$ \\
			$Z$ & Electricity cost of all workflows \\ 
			$T^B_{w,i}$ & Start time of task $v^w_i$ \\
			$T^F_{w,i}$ & Finish time of task $v^w_i$ \\
			$x_{w,i;k,j,l}$ & Binary variable for task $v^w_i$ is allocated to $\mc{V}^k_{j,l}$ \\ 
			\hline
		\end{tabular}
	\end{table}

\vspace{-2mm}

 \vspace{-2mm}
\subsection{Problem Formulation}
 \vspace{-1mm}

Each application $G_w\in {G}$ has a submission time $T^{submit}_w$ with  a user-specified deadline $d_w$. Let $T^B_{w,i}$ be the start time of $v^w_i$, and $T^F_{w,i}$ be the finish time of $v^w_i$. Each task $v^w_i$ can only start executing after all  predecessor tasks $v^w_{i^{'}}\in PR^w_i$ have finished, which should satisfy the constraint:
	\begin{equation}
		T^B_{w,i}\geq{T^F_{w,i'}},{i' \in PR^w_i}
        \label{cons_btime}
	\end{equation}
    
 In this paper, we consider multiple workflow applications with deadline constraints which arrive randomly and the heterogeneity of data center resources with  dynamic electricity prices. To describe  the problem clearly, we have the following assumptions: (i) The 
server will not fail during task execution. (ii) The task is  processed only on the assigned 
 server.
(iii) There is no performance fluctuation when the 
 server executes the task, and its processing frequency and power remain constant.
(iv) 
  The bandwidth of different data centers, clusters and servers are randomly refreshed 
  in each time step. 
(v) 
The energy is consumed by 
servers 
during task execution and data transmission.

Workflows are scheduled on different servers 
of 
clusters in GDCs. The binary decision variable $x_{w,i;k,j,l}$ is used to denote whether an arbitrary task $v^w_i$ is scheduled  to  the server $\mc{V}^k_{j,l}$ or not.
Since a task is  executed by one 
server, for each task $v^w_i$, it satisfies:
$\sum_{k=1}^M\sum_{j=1}^{\mu_k}\sum_{l=1}^{\omega^k_j} x_{w,i;k,j,l}=1 \label{cons_varx_ontask}$.
For each 
server $\mc{V}^k_{j,l}$, any two tasks assigned to it, $v^w_i$ and $v^{w'}_{i'}$, should have non-overlapping execution time with the following constraint:
	$(x_{w,i;k,j,l}\times T^B_{w,i} \ge x_{w',i';k,j,l}\times T^F_{w',i'}) \lor (x_{w,i;k,j,l}\times T^F_{w,i} \le x_{w',i';k,j,l}\times T^B_{w',i'})$.
The finish time of a task, $T^F_{w,i}$, is determined by the start time, $T^B_{w,i}$, and the execution time of task $v^w_i$ on the 
server $T^D_{w,i}$, which includes the maximum transmission time  from the predecessor task to the current task and the execution time, $T^{work}_{w,i}$, to complete the workload $W_{w,i}$:
\begin{equation}
    	T^F_{w,i}=T^B_{w,i}+T^D_{w,i}
     \label{task_finish_time} 
\end{equation}
where
   $T^D_{w,i}=T^{work}_{w,i}+\max_{{v^w_{i'}\in PR^w_i}}{{T^{trans}_{w,i',i}}}
    \label{task_duration_time} $ and
  $T^{work}_{w,i}=\sum_{k=1}^{M}\sum_{j=1}^{\mu_k}\sum_{l=1}^{\omega^k_j}{x_{w,i;k,j,l}\times \frac{W_{w,i}}{\xi_{k,j,l}}} 
      \label{task_work_time} $.

The workload $W_{w,i}$ is measured in Million of Instructions (MI), and the processing frequency of 
server $\xi_{k,j,l}$ is in MIPS (MI Per Second). The transmission time for each predecessor task $v^w_{i',i}\in PR^w_i$  to $v^w_i$ is $T^{trans}_{w,i',i}$. When tasks \( v^w_{i'} \) and \( v^w_i \) are assigned to the same 
cluster, the transmission time is 
zero 
because data transfers locally. Otherwise, the transmission time is calculated by the bandwidth \( B_{in} \) for different clusters within the same data center or \( B_{out} \) for transmission between different 
centers.
We introduce a ternary auxiliary variable $e$ to represent the different data transmission. The transmission time  $T^{trans}_{w,i',i}$ is calculated by:
	\begin{gather}
T^{trans}_{w,i',i}=\frac{e_{i',i}S_{w,i',i}}{2(e_{i',i}-1)B_{out}+(2-e_{i',i})B_{in}} \\
 \label{task_trans_time} 
 e_{i',i} = \begin{cases}
    0, &\text{if 
   servers of  $v^w_{i'}$ and $v^w_i$  are in the same 
   cluster};\\
    1, &\text{if 
    servers are in the same DC but different 
    clusters;}\\
    2, &\text{if 
    servers are in different DCs.}\\
\end{cases}	  
\nonumber
	\end{gather}
    
 The finish time of the latest completed   task of workflow $G_w$ should not exceed $d_w$:
		$\max_{{i\in \left\{1,\dots,n_w\right\}}}\left\{T^F_{w,i}\right\} \le d_w$.
	
The objective is to minimize the electricity cost  for workflow applications, which is closely related to the electricity price $p_k(t)$ of geo-distributed DCs, the execution time $T^D_{w,i}$ of the task on the 
server, and the power consumption $P_{k,j,l}$ of the
server. The mathematical model is described as follows:
	\begin{equation}
	    \min ~ Z=\sum_{w=1}^N\sum_{i=1}^{n_w}{C_{w,i}} \label{obj}
	\end{equation}
		\begin{equation}
		   C_{w,i}=\sum_{k=1}^{M}\sum_{j=1}^{\mu_k}\sum_{l=1}^{\omega^k_j}\int_{T^B_{w,i}}^{T^F_{w,i}}{x_{w,i;k,j,l}\times P_{k,j,l}\times p_k(t) dt}
     \label{task_cost} 
		\end{equation}
		 \begin{equation}
		     \min_{{i=1,\dots,n_w}}\left\{T^B_{w,i}\right\} \ge T^{submit}_{w} \label{cons_btime2} 
		 \end{equation}
		\begin{equation}
		    \max_{{i=1,\dots,n_w}}\left\{T^F_{w,i}\right\}\le d_w \label{cons_ftime}
		\end{equation}
	
\vspace{-2mm}
\section{Electricity Cost-aware  Workflow Scheduling }\label{sec4}
To minimize the electricity cost from geo-distributed DCs, we propose the Electricity Cost-aware Multiple Workflows Scheduling (ECMWS) algorithm, which considers 
servers with fixed frequency and power similar as studies \cite{8954766,10.1007/s11227-021-03733-4,Garg2020EnergyAR,10.1109/TNSM.2016.2554143}. The framework of ECMWS is shown in  Algorithm \ref{algo:1} with termination time and scheduling intervals.
Firstly, for various requirements of distinct workflows,  to avoid  repeated calculation with un-satisfied 
servers and improve algorithm efficiency, the performance parameters of 
servers such as the average processing frequency and the transmission bandwidth are pre-estimated. 
Secondly, due to the imperative to ensure the completion of all workflows prior to their respective deadlines, it is essential to implement a scheduling mechanism that prioritizes and sequences the workflows. ECMWS collects workflows  submitted to the system, estimates the execution and transmission time of tasks in workflows, and we propose the Contention aware Workflow Sequencing (CWS) algorithm to get workflow scheduling sequence in Algorihtm \ref{algo:2}.
Thirdly, given the interdependent relationship among tasks within workflows, it is essential to effectively allocate resources by implementing a prioritization and sequencing process for  tasks as well.
Thus,
for each workflow in the scheduling sequence, the Bottleneck-Layer-aware sub-Deadline Partitioning (BLDP) and Task Sequencing (TS) procedures are proposed to obtain the scheduling sequence of tasks. For resource allocation, to mitigate the overall electricity cost associated with task execution while avoiding entrapment in local optima, we 
train 
an electricity cost-aware policy network. 
For each task in the task scheduling sequence, ECMWS performs a Confidence Constrained Resource Allocation (CCRA) algorithm to allocate DCs and 
servers to it by graph embedding and Proximal Policy
Optimization (PPO). By repeating 
the 
above procedures at fixed scheduling intervals, the electricity cost is minimized.

   \begin{algorithm}[!tb]
   \small
   	\caption{ECMWS Framework}
   	\label{algo:1}
   	\KwIn{System termination time $T^{term}$, interval $\tau$}
   	\KwOut{Electricity cost $Z$}
   	
   		$Z\gets 0$, $t\gets 0$;\\
   		 Estimate VM performance parameters;\\
   		\While{$t<T^{term}$}{
   			$t'\gets \min{\{t+\tau, T^{term}\}}$;\\
   			$G\gets \emptyset$;\\
   			$G \gets G\cup workflows$ submitted  between $t$ to $t'$;\\
   			\If{${G}$$\neq\emptyset$}{
   			Estimate time parameters of  tasks for workflows;\\
   			$(G_1,G_2,\dots,G_{N'})\gets CWS(G)$;\\
   				\For{$w=1$ to $N'$}{
   					$\{(v^w_1, d_{w,1}),\dots,(v^w_{n_w},d_{w,n_w})\}\gets BLDP(G_w)$ .\\
   					$(v^w_1,v^w_2,\dots,v^w_{n_w})\gets TS(G_w)$;\\
   					\For{$i=1$ to $n_w$}{
   						$({D}_k, {V}^k_{j,l})\gets CCRA(v^w_i)$;\\
                          Compute $C_{w,i}$ by Eq.  \eqref{task_cost}.\\
   						$Z\gets Z + C_{w,i}$;\\
   					} 
   				}
   			}
   			$t\gets t'$;\
   		}
   \end{algorithm}

To illustrate the ECMWS framework, we assume that there are two workflows to be allocated: workflow 1 with 17 tasks and  workflow 2 with 6 tasks with 1 hour 
deadline. 
We follow the common setting in data center platforms, like Cisco Nexus 5600\footnote{https://www.cisco.com/c/en/us/products/collateral/switches/nexus-5624q-switch/datasheet-c78-733100.html}.
    There are three cross-domain data centers with electricity price (0.16\$ all day, and 0.16\$ (0:00-8:00), 0.19\$ (8:01-23:59)). Each data center has two clusters, and each cluster has two servers.
    The processing frequencies  of servers
    are  1,000 MIPS
    and  500 MIPS,
     respectively. To better simulate the actual network scheduling, we set up  dynamic bandwidths for different states. 
When the network is in a congested state, the actual bandwidth would be randomly reduced \cite{9339423} (30\%-70\% of the corresponding bandwidth in this paper).
The bandwidth between data centers is randomly changed between 0 to 100 Gbps,
      while the bandwidth between 
     clusters
     in a data center is set to   
     0 to 80 Gbps.
    All data transmission between 
  servers
    within the same 
 cluster
    have a 
    consistent
   random bandwidth 
    between 0 to 40 Gbps.
    The specific scheduling process with the same example is discussed in the following subsections. 


\vspace{-4mm}
\subsection{Parameter Estimation}\label{sec_p}
\vspace{-1mm}
For geo-distributed DCs, there are exponential growth combinations for multiple workflows with heterogeneous servers.
To avoid redundant computations and enhance the efficiency, ECMWS estimates the  parameters of 
servers during the initialization phase including the average processing frequency of 
servers and the average data transmission bandwidth between 
servers. 
The former is defined as the average processing frequency of all 
servers in  cross-domain DCs where
$\hat{\xi}=\frac{\sum_{k=1}^M\sum_{j=1}^{\mu_k}\sum_{l=1}^{\omega^k_j}\xi_{k,j,l}}{\overline{\omega}} \label{vm_avg_speed}$.
The latter is  the average of the combined transmission bandwidth of all pairwise virtual groups:
\begin{equation}
\begin{aligned}
  	\overline{B}=&\frac{B_{in}\times \sum_{k=1}^M\sum_{j=1}^{\mu_k-1}\sum_{j'=j+1}^{\mu_k}(\omega^k_j\times \omega^k_{j'})}
   {\overline{\omega}^2}
   \\&+\frac{B_{out}\times \sum_{k=1}^{M-1}\sum_{k'=k+1}^M(\omega^k\times \omega^{k'})}{\overline{\omega}^2} \label{vm_avg_bandwidth} 
   \end{aligned}
\end{equation}
  where  $ \omega^k=\sum_{j=1}^{\mu_k}\omega^k_j \label{vm_per_dc}$.
Given a task $v^w_i$, since the task may be assigned to any 
server, the average processing frequency of 
server $\overline{\xi}$ is used to estimate the actual execution time of $v^w_i$:
\begin{equation}
\hat{T}^{work}_{w,i}=\frac{W_{w,i}}{\hat{\xi}}   \label{task_avg_worktime}
	\end{equation}

The average data transfer bandwidth $\overline{B}$ between 
servers $\hat{T}^{trans}_{w,i}=\max_{v^w_{i'}\in PR^w_i}\left\{\hat{T}^{trans}_{w,i',i}\right\} 
\label{task_avg_transtime}$ is used to estimate the data transmission time of the task,
\begin{equation}
      \hat{T}^{trans}_{w,i', i}=\frac{S^w_{i',i}}{\overline{B}}
      \label{taskpair_avg_transtime}
\end{equation}
where $\hat{T}^{trans}_{w,i',i}$ is the data transmission time of task $v^w_i$ and its immediate predecessor task $v^w_{i'}$.

According to the task execution time and data transmission time,  the earliest start time $\hat{T}^{EST}_{w,i}$ and the earliest finish time  $\hat{T}^{EFT}_{w,i}$ of each task is computed  by a 
dynamic programming method\cite{1966Dynamic} based on the execution time $\hat{T}^{work}_{w,i}$ and data transmission time $\hat{T}^{trans}_{w,i}$.
	\begin{equation}
 		\hat{T}^{EFT}_{w,i}=\hat{T}^{EST}_{w,i} + \hat{T}^{trans}_{w,i} + \hat{T}^{work}_{w,i} \label{task_avg_eft} 
 	\end{equation}
  \begin{equation}
      \hat{T}^{EST}_{w,i}= \begin{cases}
 			\lceil \frac{T^{submit}_w}{\tau} \rceil\times \tau, & \text{if } PR^w_i=\emptyset \\
 			\max_{v^w_{i'}\in PR^w_i}\left\{\hat{T}^{EFT}_{w,i'}\right\}, & \text{else} 
    \label{task_avg_est} 
 		\end{cases}
  \end{equation}

According to Eq.  \eqref{cons_btime}, Eq.  \eqref{task_avg_est} indicates that the earliest start time of each task is determined by 
its direct predecessor tasks with the latest completion time.
 The parameter estimation consists of two parts: {\em initialization} and {\em scheduling} in Algorithm \ref{algo:1}. 
 For all $N$ workflows submitted to the system, the 
 time complexity of parameter estimation is $O(\overline{\omega}^2 + \sum_{w=1}^N n_w^2)$. Parameter estimation is essential for assessing workflow workloads and reducing the resource waste generated by un-suitable 
servers. 
In multiple workflow environments, it guides the order of execution of tasks that can be used in resource allocation to improve overall efficiency. 

\vspace{-2mm}
\subsection{Workflow Sequencing}
 \vspace{-1mm}

Optimal scheduling results are achieved if the system has abundant resources while resources are usually limited, which leads to local optimal when optimizing electricity costs for multiple workflow allocations. 
Interleaved scheduling struggles to meet requirements and constraints within limited computational complexity. 
Therefore, a reasonable workflow sorting strategy is applied based on their contention where the 
CWS algorithm is proposed.
CWS 
determines the workflow sequence defined by the workload factor $WL(G_w)$, slack time $ST(G_w)$, and resource contention factor $CT(G_w)$ for $G_w$. 
$WL(G_w)$ is the normalized value of the sum of all workflow tasks $W_{w}$:
\begin{equation}
		WL(G_w)=\frac{W_{w}}{\max_{w'=1,\dots,N'}{W_{w'}}} \label{factor_workload} 
  \end{equation}
		where $W_{w}=\sum_{i=1}^{n_w}W_{w,i} \label{workflow_workload}$.
The slack time factor $ST(G_w)$ is denoted as the normalized value of the workflow slack time:
\begin{equation}
ST(G_w)=\frac{T^{slack}_w}{\max_{w'=1,\dots,N'}{T^{slack}_{w'}}} \label{factor_slack_time}
\end{equation}
where the slack time $T^{slack}_{w}$ is the difference between the earliest completion time $\hat{T}^{EFT}_{w}$ and  deadline $d_w$:
$T^{slack}_{w}=d_w-\hat{T}^{EFT}_{w} \label{workflow_slack_time}$.
The smaller slack time indicates that the application is more urgent.
$\hat{T}^{EFT}_{w}$ is determined by the completion time of last task on its critical path:
$\hat{T}^{EFT}_w=\max_{i=1,\dots,n_w}{\hat{T}^{EFT}_{w,i}} \label{workflow_avg_eft}$.
The resource contention factor $CT(G_w)$ is  the normalized value of the workflow resource contention ${Contention}_w$:
   	\begin{equation}
   		CT(G_w)=\frac{{Contention}_{w}}{\max_{w'=1,\dots,N'}{{Contention}_{w}}} \label{factor_contention}
     \end{equation} 
where ${Contention}_w$ represents the maximum number of 
servers that the workflow may occupy simultaneously. 

To calculate  CWS, the procedures are described in Algorithm \ref{algo:2}. Firstly, the execution time interval of each task from $\hat{T}^{EST}_{w,i}$ to $\hat{T}^{EFT}_{w,i}$  is added to the queue $Interval$  in ascending order of $\hat{T}^{EST}_{w,i}$.
The number of overlapping time intervals $curContention$ of a task with  other later tasks  is counted by traversing $Interval$ with maximum value $Contention_w$.
The $ComputeContention$ function is used to compute the resource contention level $Contention_w$ and the contention factor $CT(G_w)$.  To balance these factors, CWS defines the rank of a workflow application $G_w$ as the weighted sum of the factors measured by parameters $\alpha_1, \alpha_2, \alpha_3\in[0,1]$:
\begin{equation}
     	Rank(G_w)=\alpha_1ST(G_w) + \alpha_2 WL(G_w) + \alpha_3 CT(G_w) \label{CWS_rank}
\end{equation}

To reduce the complexity of parameter correction by shrinking the parameter space,  
the following constraint should be satisfied without losing generality:
\begin{equation}
    \alpha_1 + \alpha_2 + \alpha_3 = 1 \label{cons_cws_alpha}.
\end{equation}

Finally, the rank $Rank_w$ is calculated, and the scheduling sequence is obtained in the ascending order of rank.
 Given a total of $T$ batches of workflows to be scheduled, the size of each batch $\left\{N_1,\dots,N_{T}\right\}$ and the constraint condition $\sum_{t=1}^{T}N_t=N$, the total time complexity of the workflow scheduling algorithm is $O(\sum_{w=1}^{N}n_w\log n_w + \sum_{t=1}^{T}N_t\log N_t)$.
\begin{algorithm}[!ht]
\small
\caption{CWS}
\KwIn{workflow set $G$,  weight parameters $\alpha_1, \alpha_2, \alpha_3$}
\KwOut{workflow scheduling sequence $Seq$}
${R}\gets\emptyset$\;
\For{$G_w\in G$}{
Calculate $WL(G_w)$ according to Eq.  \eqref{factor_workload} \;
Calculate $ST(G_w)$ according to Eq.  \eqref{factor_slack_time}\;
${Contetion}_w\gets \text{ComputeContention}(G_w)$ \;
Calculate $CT(G_w)$ according to Eq.  \eqref{factor_contention}\;
Calculate ${Rank}_w$ by Eq.  \eqref{CWS_rank};\\
$R\gets R\cup (G_w, {Rank}_w)$ \;
}
$Seq \gets \text{Sort workflows } $in $G$ ascendingly based on $R$\;
\label{algo:2}
\end{algorithm}
For example, when $\alpha_1=0.4, \alpha_2=0.2$ and $\alpha_3=0.4$, according to Equations \eqref{factor_workload}, \eqref{factor_slack_time}, \eqref{factor_contention}, these parameters are calculated as follows: $WL(G_1):0.72, WL(G_2):0.28, ST(G_1):0.30, ST(G_2):0.16, CT(G_1):1, CT(G_2):1$. Thus, the  workflow scheduling sequence is computed by Eq.  \eqref{CWS_rank} with $Seq = \{G_1,G_2\}$. 

\vspace{-2mm}
\subsection{Deadline Partition}
 \vspace{-1mm}

To improve the efficiency and fairness of workflow allocation, workflow deadlines should be refined into sub-deadlines of each task. By appropriately assigning sub-deadlines to each task,  all tasks are completed within the partition deadline meanwhile satisfying the  workflow deadline constraint~\cite{7887706}. With limited resources, if the sub-deadlines of tasks are too wide or too tight,  it will lead to low resource utilization and long completion time of workflow which may violate the deadline constraints.
 To comprehend the dependency  among tasks in the workflow, tasks are classified into various levels, where tasks within the same level can be executed in parallel, while tasks with different levels exhibit  dependency relationship. 
  We propose a 
  BLDP strategy to partition sub-deadlines for a workflow by calculating the rank of tasks. Firstly, it calculates  task ranks by the upstream completion time \cite{10.1109/71.993206}. 
  Secondly, BLDP partitions each task $v^w_i$ in  $G_w$ into a level $l_{w,i}$ based on its dependencies and the number of tasks $\phi(l_{w,i})$ at the same level  in a workflow. 
    	$\phi(l_{w,i})=|\left\{v^w_j | l_{w,j}=l_{w,i}\right\}| $ 
  where  $l_{w,i}=1$ if $SU^w_i=\emptyset$, otherwise $l_{w,i}=\max_{v^w_j\in SU^w_i}{l_{w,i}}+1$.
To compute the bottleneck-layer-aware rank $Rank^{DP}_{w,i}$ for $v^w_i$, we use the estimated task processing time $\hat{T}^{work}_{w,i}$ by Eq.  \eqref{task_avg_worktime}, and the estimated data transmission time $T^{trans}_{w,i,j}$ by Eq.  \eqref{taskpair_avg_transtime}. The rank is calculated by:
\begin{equation}
		Rank^{DP}_{w,i}= \begin{cases}
			\hat{T}^{work}_{w,i}, & \text{if } SU^w_i=\emptyset \\
     
					\max_{v^w_j\in SU^w_i}(Rank^{DP}_{w,j}+\hat{T}^{trans}_{w,i,j}) \\+ \hat{T}^{work}_{w,i} +
     \beta^{\frac{\phi(l_{w,i})}{\phi(l_{w,i}-1)}}, & \text{else } \label{rank_DP}
				\end{cases}
	\end{equation}

Finally, the sub-deadline $d_{w,i}$  for each task $v^w_i$ is calculated by its rank and the workflow deadline $d_w$.
\begin{equation}
    d_{w,i}=d_w\times\frac{Rank^{DP}_{w,0}-Rank^{DP}_{w,i}+\hat{T}^{work}_{w,i}}{Rank^{DP}_{w,0}} \label{task_deadline}
\end{equation}
\begin{equation}
    Rank^{DP}_{w,0}=\max_{\substack{1\leq i \leq n_w \\ PR^w_i=\emptyset}}{Rank^{DP}_{w,i}} \label{rank_DP_0}
\end{equation}

Based on  above analysis, the sub-deadline for each task of  workflow 1 is calculated as follows: $d_{1,1}=0.12, d_{1,2}=0.16, d_{1,3}=0.23,\cdots,d_{1,17}=1$.

 \vspace{-2mm}
\subsection{Task Sequencing}
 \vspace{-1mm}

The task scheduling sequence has a significant impact on the electricity cost incurred during the task scheduling process. Given a workflow $G_w$ and a task $v^w_i\in G_w$, the upward rank $Rank^{up}_{w,i}$ and downward rank $Rank^{down}_{w,i}$ are calculated by the upstream and downstream completion times \cite{10.1109/71.993206}, respectively. 
We construct the task scheduling sequence by sorting $v^w_i$ in descending order according to $Rank^{up}_{w,i}$ or in ascending order according to $Rank^{down}_{w,i}$. The definitions of the two types of task rank are described as follows:
\begin{equation}
Rank^{up}_{w,i}=\max_{v^w_j\in SU^w_i}{\hat{T}^{trans}_{w,i,j} + Rank^{up}_{w,j}} + \hat{T}^{work}_{w,i}  \label{rank_up}
\end{equation}
\begin{equation}
Rank^{down}_{w,i}=\begin{cases}
0, & \text{if }PR^w_i=\emptyset \\
\max_{v^w_j\in PR^w_i}(\hat{T}^{trans}_{w,j,i}\\+\hat{T}^{work}_{w,j}+Rank^{down}_{w,i}), & \text{else}
\end{cases} \label{rank_down}
\end{equation}

We reuse the  rank $Rank^{DP}_{w,i}$ 
by Eq.  \eqref{rank_DP} to ensure that tasks with higher ranks are executed earlier than tasks with lower ranks. Tasks are sorted 
as follows:
(1) $TS_1$: Calculate the upward rank $Rank^{up}_{w,i}$ for each task $v^w_i$ by Eq.  \eqref{rank_up} and sort  in decreasing order.
(2) $TS_2$: Calculate the downward rank $Rank^{down}_{w,i}$ for each task $v^w_i$ by Eq.  \eqref{rank_down} and sort
in increasing order.
(3) $TS_3$: Calculate the task rank $Rank^{DP}_{w,i}$ for each task $v^w_i$ by Eq.  \eqref{rank_DP} and sort 
in decreasing order.

 Since computing $Rank^{up}_{w,i}$ and $Rank^{down}_{w,i}$ requires examining every combination of a task with all direct predecessors or successors, the total time complexity of $TS_1$ and $TS_2$ for $N$ workflows is 
 $O(\sum_{w=1}^N n_w^2)$ while the 
 time complexity of $TS_3$ is $O(\sum_{w=1}^N n_w\log n_w)$, which only requires sorting the workflow tasks based on the task rank values computed in the previous stage.
According to $TS_3$, the task sequence of 
workflow 1 is $ (v^1_1, v^1_2, v^1_4, v^1_3,v^1_5, v^1_6,v^1_7, v^1_9,v^1_8, v^1_{10},v^1_{11}, v^1_{12},v^1_{13}, v^1_{15},v^1_{14}, v^1_{17}$ $,v^1_{16})$, and 
for workflow 2, it is $ (v^2_6, v^2_5, v^2_2, v^2_4,v^2_3, v^2_1) $.

\vspace{-2mm}
\subsection{Resource Allocation}
For task resource allocation, the objective is to allocate 
servers effectively to reduce the electricity cost of task execution while satisfying the task sub-deadline constraints. However, solely minimizing the electricity cost incurred by individual tasks can lead to local optimal optimization of the overall electricity cost for all workflows. Therefore, 
the Proximal Policy Optimization (PPO) algorithm \cite{schulman2017proximal}  
is 
employed 
to train an electricity-cost-aware policy network. 
This policy network takes a global view of the batch workflow scheduling process and dynamically assigns each task to the optimal data center and the specific 
server based on the real-time system state, approximating the overall scheduling objective optimally. Subsequently, a 
CCRA algorithm is proposed based on this policy network, which incorporates confidence constraints into the task resource allocation process.
In Algorithm \ref{algo:4}, the task and resource graph embedding models are constructed to generate the state vectors, respectively. These vectors are used as input to call the policy network to generate the probability distribution of resource allocation actions, and the action with the highest probability is selected. If the probability of the selected action is no less than a given threshold, the DC and 
server selected by this action are adopted. Otherwise, a fallback policy is executed to select a new DC and 
server.
\begin{algorithm}[!ht]
\small
\caption{CCRA}
\label{algo:4}
\KwIn{Task $v^w_i$; Workflow $G_w$;  Confidence threshold $Conf_{thresh}$}
\KwOut{DC $D_k$ and 
server $V^k_{j,l}$ assigned to $v^w_i$}
Construct 
$model_t$ and 
$model_r$;\\
$\Vec{S}_{task}\gets model_t(v^w_i, G_w)$;\\
$\vec{S}_{
server}\gets model_r({D_1,\dots,D_M})$;\\
Obtain Policy network $\pi_{\theta}$ by RAPPO in Algorithm \ref{algo:ppo_training};\\
Construct the price state vector $\vec{S}_{price}$ by Eq.  \eqref{vec_price};\\
Construct  $\vec{S}_{params}$ by Eq.  \eqref{vec_params};\\
$\vec{S}_t\gets \begin{bmatrix}
\vec{S}_{task} & \vec{S}_{
server} & \vec{S}_{price} & \vec{S}_{params}
\end{bmatrix}$; \\

$Pr_{A|S}\gets \pi_{\theta}(\vec{S}_t)$;\\
$A_t \gets \arg\max_{A}\left\{Pr_{A|\vec{S}_t}\right\}$;\\
$Conf_t \gets \max_{A}\left\{Pr_{A|\vec{S}_t}\right\}$;\\
\If{$Conf_t\ge Conf_{thresh}$}{
$(D_k, V^k_{j,l})\gets A_t$ ;\\
}
\Else{
$(D_k, V^k_{j,l})\gets DARA(v^w_i)$ ;
}
\end{algorithm}

For workflows $G=\{G_1,\dots,G_N\}$, the scheduling process involves decision points of tasks assignment. The system state $S_t\in \mathcal{S}$ at time step $t$ is considered as input, and the policy network is trained to generate resource allocation actions $A_t \in \mathcal{A}$. By constructing Markov Decision Process (MDP) that encompasses $\overline{n}=\sum\limits_{G_w\in G}n_w$ time steps for geo-distributed DCs, the electricity cost is  minimized  with multiple workflows. Each component of the MDP $\mathcal{M}=(\mathcal{S},\mathcal{A},\mathcal{T},\mathcal{R},\gamma,\rho_0)$ is described as follows.
   
    $\mathcal{S}$: a continuous space, where each state $S_t\in \mathcal{S}$ is constructed by concatenating the task embedding vector $\vec{S}_{task}$, resource embedding vector $\vec{S}_{
   server}$, electricity price vector for cross-domain DCs $\vec{S}_{price}$, and parameter vector of the algorithm $\vec{S}_{params}$. 
    $\vec{S}_{price}$ is constructed by concatenating the normalized values of the average electricity prices for each hour in  future, $Dim_{price} $ at each DC. These prices are calculated by the current hour $h$ (represented in UTC time) of the cross-domain DC system:
    \begin{equation}
			\vec{S}_{price}=\frac{\vec{p}-\vec{p}_{min}}{\vec{p}_{max}-\vec{p}_{min}} \label{vec_price} \\
   \end{equation}
   where 
			$\vec{p}=\begin{bmatrix}		\hat{p}_1(h),\dots,\hat{p}_1(h+Dim_{price}-1),\dots,\\
   \hat{p}_M(h),\dots,\hat{p}_M(h+Dim_{price}-1)
			\end{bmatrix}$,	 	
			$\vec{p}_{min}=\begin{bmatrix}
				\multicolumn{3}{c}{\overbrace{p_{min}\quad\cdots\quad p_{min}}^{D_{price}}}\end{bmatrix}$,
			$\vec{p}_{max}=\begin{bmatrix}
				\multicolumn{3}{c}{\overbrace{p_{max} \quad \cdots \quad p_{max}}^{D_{price}}}\end{bmatrix}$, 
 $p_{min}=\min_{\substack{k=1,\dots,M \\ t=0,\dots,24\times 3600}}{p_k(t)}$, 
         $p_{max}=\max_{\substack{k=1,\dots,M \\ t=0,\dots,24\times 3600}}{p_k(t)}$, 
      and $\hat{p}_k(h)=\frac{\sum_{t=h\times 3600}^{(h+1)\times 3600}p_k(t)}{3600}$.
$\vec{S}_{params}$ is embedded into the State Space  by CWS, BLDP, and Task sorting Parameters:
\begin{equation}
\vec{S}_{params}=\begin{bmatrix}
\overbrace{\alpha_1\quad\alpha_2\quad\alpha_3}^{\text{CWS parameters}} & \overbrace{\beta}^{\text{BLDP parameter}} & \overbrace{0\quad0\quad 1}^{\text{Task sorting }}
\end{bmatrix} \label{vec_params}
\end{equation}
    
 $\mathcal{A}$: For a discrete action space,  each action ($A_t \in \mathcal{A}$) denotes the allocation of a DC and a 
server to a task. $\mathcal{A}$ is the Cartesian product of two discrete sub-action spaces:
$\mathcal{A} = \mathcal{A}^{dc}_M \times \mathcal{A}^{
server}_{\omega_{max}} \label{factored_action}$.
 $\mathcal{A}^{dc}_M$ represents the action space for selecting a DC where $M$ is the number of DCs. $\mathcal{A}^{
server}$ represents the action space for selecting a 
server, where $\omega_{max} = \max_{k\in{1,\dots,M}}\left(\sum_{j=1}^{\mu_k}\omega^k_j\right)$ is the maximum number of 
servers across all DCs. The policy network  samples from these two action spaces  to generate sub-actions for selecting a DC and selecting a 
 server, respectively.

$\mathcal{T}: \mathcal{S} \times \mathcal{A} \times \mathcal{S} \rightarrow [0,1]$ is the state transition function, which provides the probability of transitioning from current state to  next state given a particular action. The PPO algorithm is employed to optimize the policy network.

$\mathcal{R}: \mathcal{S} \times \mathcal{A} \times \mathcal{S} \rightarrow \mathbb{R}$ is the reward function with real-valued outputs. Based on the optimization objective and the deadline constraints of the workflow, the following reward function is designed for the task $v^w_i$ to be scheduled at time step $t$:
\begin{equation}
R_t = -C_{w,i} \times \left(1 + \max\left(0, \frac{T^F_{w,i} - d_{w,i}}{T^F_{w,i} - T^B_{w,i}}\right)\right) \label{reward}
\end{equation}

The task's electricity cost $C_{w,i}$ and finishing time $T^F_{w,i}$ are computed by Equations \eqref{task_cost} and \eqref{task_finish_time}, respectively.  The objective of $\mathcal{M}$ is to learn the optimal policy network parameters $\theta$.
At each time step $t$, the policy network $\pi_{\theta}$ is utilized to select the best DC and 
 server for state $s$, thereby maximizing the expected cumulative reward $V^{\pi}(s)$:
$V^{\pi}(s) = \mathbb{E}_{\pi}\left[G_t \mid S_t = s\right]$.

$\gamma$: The discount factor is used to calculate the discounted cumulative reward $G_t$, which evaluates the long-term value of taking an action at a particular time step:
\begin{equation}
G_t = \sum_{i=0}^{\overline{n}-t-1} \gamma^i R_{t+i}. \label{acc_return}
\end{equation}

$\rho_0$: initial state distribution. For the scheduling process of different batch workflows, the characteristics of the first task of workflows to be scheduled and the availability of 
 server resources  varies, resulting in different initial states of $\mathcal{M}$.

\subsubsection{Graph Embedding Model for Tasks and Resources}
\begin{figure}[tbp]
		\centering
		\includegraphics[width=0.48\textwidth]{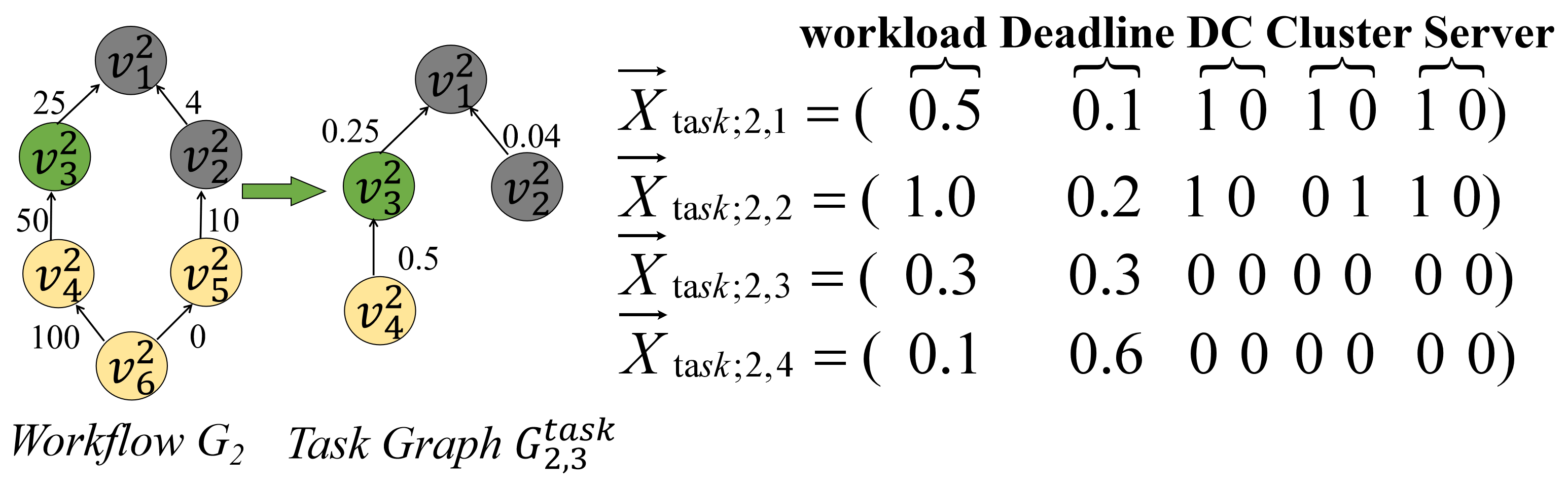}
		\caption{Task diagram example.}
		\label{fig:Task_graph}
  \vspace{-7mm}
	\end{figure}

According to the partial order relation constraint of tasks in workflows  in Eq.  \eqref{cons_btime}, the structure of workflow is crucial to the scheduling of workflow tasks. 
Graph Convolutional Network (GCN) only considers the local neighborhood information of the node, which enables the model to capture the local connection pattern of the node more efficiently. 
By employing a graph embedding model $model_t$  with two-layer GCN, the high-dimensional structured workflow task information is encoded into a low-dimensional vector, where the output dimensions  are $D_{tg}$, $D_{tn}$ in the Encoder, and $D_{tg}$, $D_{task}$ in the Decoder.
Fig. \ref{fig:Task_graph} shows an example for
workflow $G_2$ and a scheduled task $v^2_{3}$ at the current time step. For the task graph, each node represents a workflow task, including task workload, deadline, and the index of the assigned 
 servers. The directed edges between nodes represent the partial order relationship between tasks, and the edge weight is the amount of data transfer between tasks. The task graph  at time step $t$ is built with the unscheduled task $v^w_i$. 
 
\begin{figure}[!tbp]
		\centering
		\includegraphics[width=0.35\textwidth]{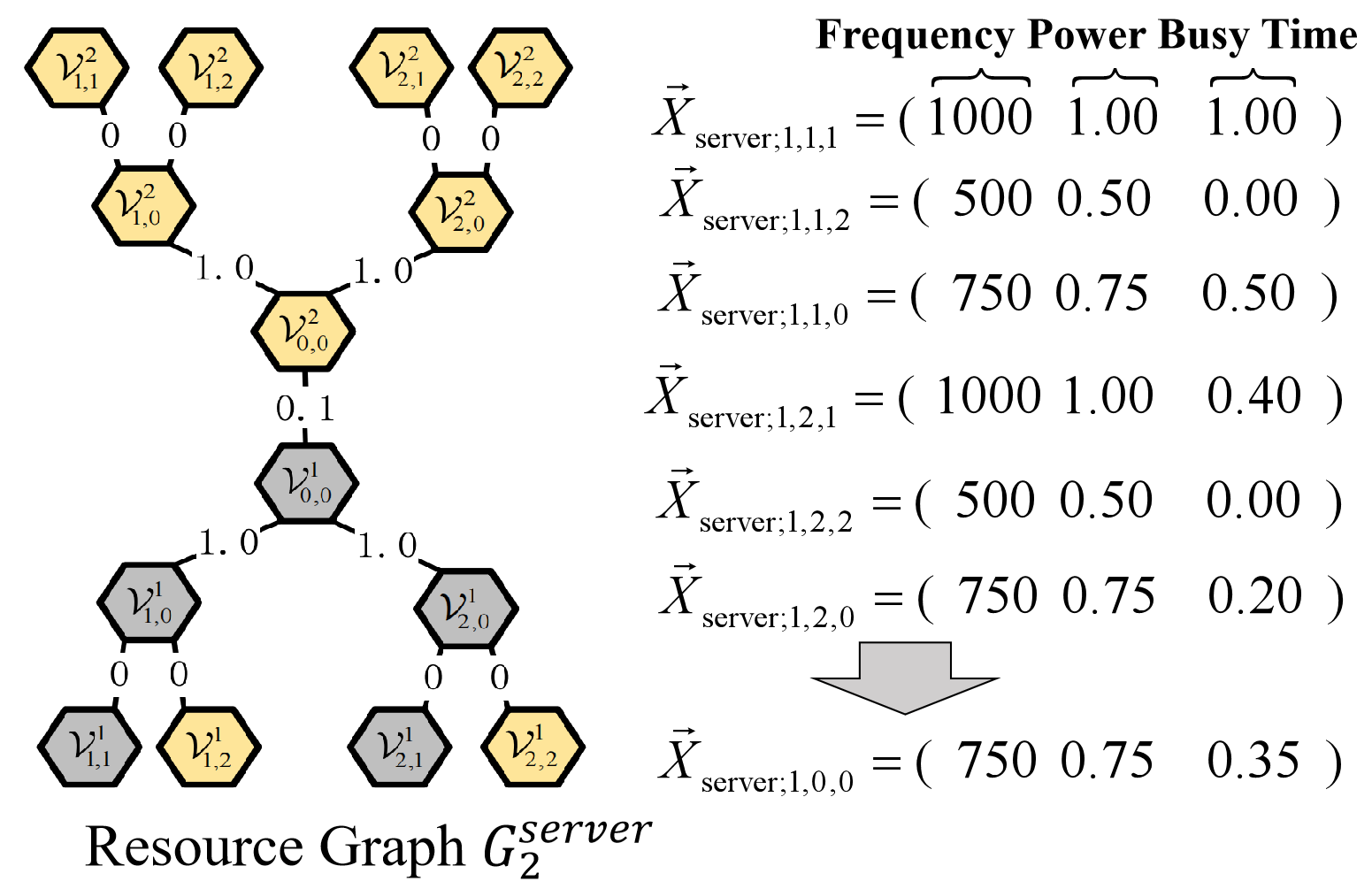}
		\caption{Resource diagram example.}
		\label{fig:resource_graph}
	  \vspace{-4mm}
\end{figure}

Similarly, the resource is  organized into an undirected graph according to their spatial relationships by the graph embedding model $model_r$.
The main difference between the two models lies in the different numbers of layers of GCNs.  To facilitate representation, we use constructed 
 server nodes $\mc{V}^k_{j,0}$ to represent 
cluster nodes $\mc{H}^k_j$, and $\mc{V}^k_{0,0}$ to represent data center nodes $\mc{D}_k$. Each resource graph $G_{t}^{
server}=(V_{t}^{
server},E_{t}^{
 server})$ at time step $t$ includes all 
 servers, 
   clusters, and data center nodes, but the features of each node vary. Therefore, the number of nodes in each resource graph $G_{t}^{
   server}$, denoted as $|V_{t}^{
server}|=\overline{\omega}+\sum_{k=1}^{M}\mu_k+M$, remains fixed.    
If a node is a 
server, its features include processing frequencies, power, and busy time; if it is a 
 cluster, the feature values are the average  features of all 
   server on the 
    cluster; if it is a DC, the feature is the average features of all 
   cluster in the DC. 
 For $model_r$
 the output dimensions of each layer in the Encoder are all $D_{rg}$, and the output dimensions of the first two layers in the Decoder are $D_{rg}$, and the output dimension of the last layer is $D_{
 server}$. 
Fig. \ref{fig:resource_graph} shows the resource diagram at time step t = 2 in the above example  with two data centers, each with two 
clusters, and each 
cluster with two 
 servers. 
 
In order to eliminate the scale difference of input data, the maximum and minimum value normalization is used to process the node features and edge weights in the task graph and resource graph to ensure  the corresponding values within the range of $[0,1]$.
For  time step $t$ and its unscheduled task $v^w_i$, given the embedding vector $\vec{X}_{task;w,i'}=(X_{task;w,i',[1]},\dots,X_{task;w,i',[D_{tn}]})$ of the task  node $v^w_{i'}\in E^{task}_{w,i}$, the task graph embedding vector $\vec{S}_{task}$ is represented by the mean of the embedding vectors of all task nodes:
\begin{equation}
\begin{aligned}
\vec{S}_{task}=\begin{bmatrix}
\frac{1}{|V^{task}_{w,i}|}\sum_{v^w_{i'}\in V^{task}_{w,i}}X_{task;w,i',[1]} & \cdots & \\ \frac{1}{|V^{task}_{w,i}|}\sum_{v^w_{i'}\in V^{task}_{w,i}}X_{task;w,i',[D_{tn}]} & \cdots & 
\end{bmatrix} \label{vec_task}
\end{aligned}
\end{equation}

At  time $t$, given the embedding vector $\Vec{X}_{
server;k,j,l}=(X_{
server;k,j,l,[1]},\dots,X_{
server;k,j,l,[D_{rn}]})$ of the resource  node $v^k_{j,l}\in E^{
 server}_{t}$, the resource graph embedding vector $\vec{S}_{
 server}$ is obtained by concatenating  embedding vectors of all DC and 
cluster nodes:

\begin{small}
\begin{equation}
\begin{bmatrix}
\overbrace{\vec{X}_{
 server;1,1,0}\cdots
 \vec{X}_{
server;M,\mu_M,0}
 }^{\sum_{k=1}^{M}\mu_k} &\overbrace{\vec{X}_{
 server;1,0,0}\cdots\vec{X}_{
server;M,0,0}}^{M}.
\end{bmatrix} \label{vec_vm}
\nonumber
\end{equation}
\end{small}

The two graph embedding models are trained by the graph autoencoder \cite{hinton2014autoencoder,SUN2021107564}. 


\subsubsection{Actor-Critic framework for workflows and resources }
To provide optimal resource allocation actions $A_t$ at  time $t$  based on its real-time state in the MDP $\mathcal{M}$, we use the Actor-Critic framework \cite{10.5555/3009657.3009806}, which combines two  reinforcement learning methods: {\em value-based} and 
{\em policy gradient-based}. 
The policy network serves as the Actor, while a value function network, Critic, evaluates the quality of actions selected by the Actor. Both the Actor and Critic networks are implemented  by a simple Multilayer Perceptron \cite{lecun2015deep}.

\begin{figure}[!tbp]
		\centering
		\includegraphics[width=0.48\textwidth]{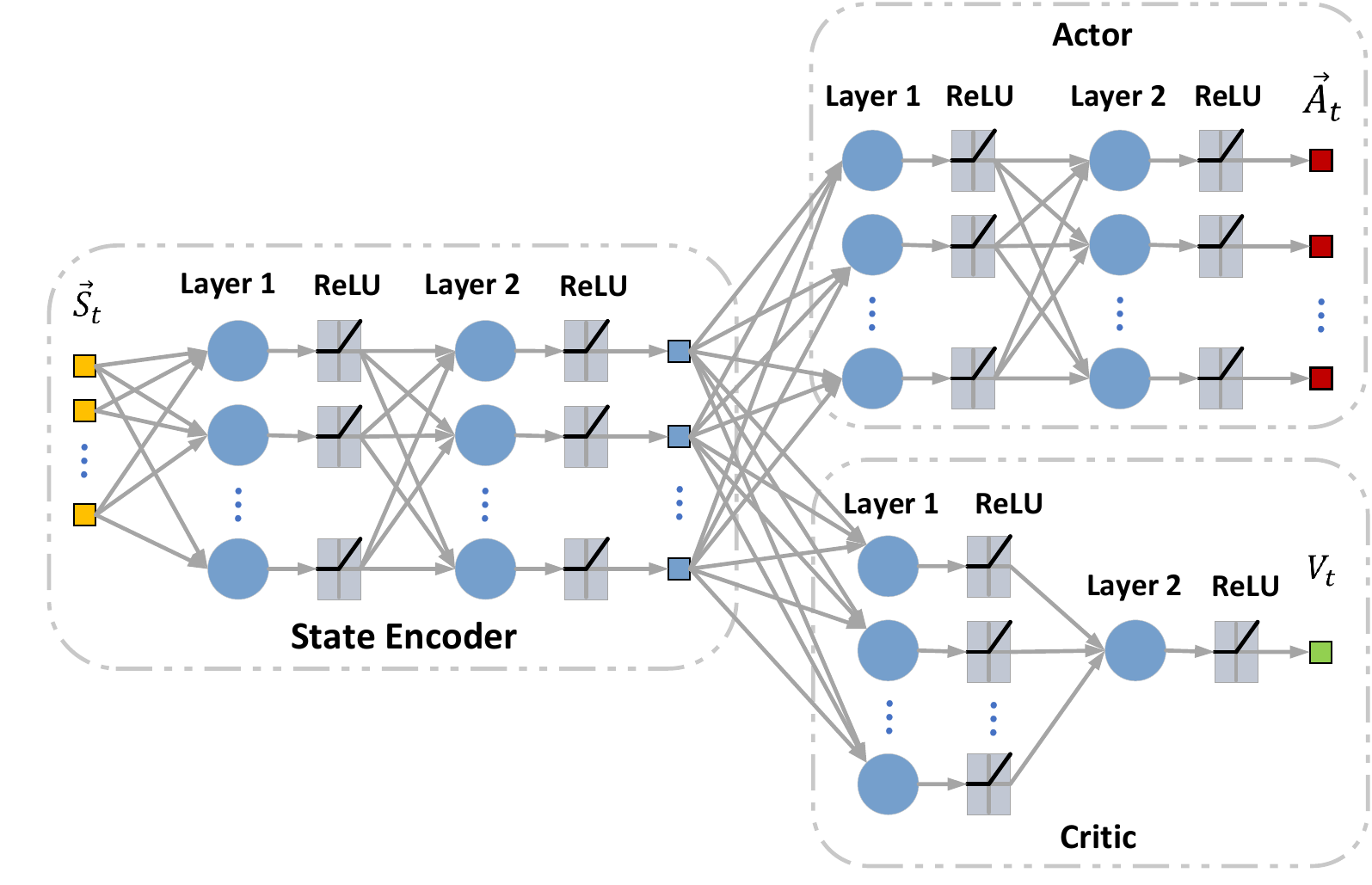}\\
		\caption{Actor-Critic network architecture.}
		\label{fig:actor_critic_arch}
  \vspace{-6mm}
	\end{figure}
 
The network structure of the Actor-Critic framework 
is described in Fig. \ref{fig:actor_critic_arch}. The two networks  share a state encoding module, which consists of two feedforward layers. This module transforms the input state vector into a hidden representation to both networks. The output dimensions of the two layers are $D_{encoder}(1)$ and $D_{encoder}(2)$, respectively. Subsequently, the Actor and Critic networks individually pass through two more feedforward layers to output $c$-dimension action vectors  and $1$-dimension action value estimation, respectively. 

To select the most suitable DC $\mathcal{D}_k$ from $M$ DCs and choose the optimal 
server from the $\omega^k$ 
 servers in the DC, the Actor network outputs an action vector representing a probability distribution over the available actions. It can be further divided into sub-action probability vectors for selecting the DC and  
 server. Sampling from these two vectors, it yields the indexes of selected DC  and 
  server. Since the output dimension of a neural network determines the number of parameters and connections, it must be fixed. When selecting the 
 server, we consider the maximum number of 
 servers $\omega_{max}$ among all DCs to compute the upper limit of the 
  server index range. We use the Action Masking technique \cite{vinyals2019grandmaster} to mask out invalid actions beyond the range of the selected DC's 
 server indexes to ensure that the dimension $D_{action}$ of the action vector can be fixed, calculated as:
$D_{action} = M + \omega_{max} \label{dim_action}$.

The action masking technique is employed to improve the efficiency and stability of the policy by limiting the action space. 
For a given sub-action of selecting DC $\mathcal{D}_k$, the sub-action masking for selecting the 
server is determined by the maximum number of 
servers in the selected DC. A masking vector $\vec{Y}_{mask}$ of dimension $\omega_{max}$ is constructed as follows, where all elements with indexes exceeding $\omega^k$ are set to 0:
\begin{equation}
\vec{Y}_{mask}=\begin{bmatrix}
\multicolumn{3}{c}{\overbrace{1 \quad \cdots \quad 1}^{\omega^k}} & \multicolumn{3}{c}{\overbrace{0 \quad \cdots \quad 0}^{\omega_{max}-\omega^k}}
\end{bmatrix} \label{vm_action_mask}
\end{equation}

The input dimension $D_{state}$ of 
the 
Actor and Critic networks is fixed and equals to the sum of dimensions of the task graph embedding vector,  resource graph embedding vector, electricity price vector, and  algorithm parameter vector:
$D_{state}=D_{task}+D_{
server}+D_{price}+D_{params}\label{dim_state}$.
Specifically, at the output layer of the policy network, the original action probability vector is multiplied element-wise with a masking vector containing 0 or 1, which restricts the number of available actions for different states. This modification only affects the value of the policy gradient and does not impact the internal structure or parameters of the policy network. Therefore,  
the PPO framework remains applicable.

\begin{algorithm}[!ht]
\small
\caption{Resource Allocation PPO (RAPPO)}
\label{algo:ppo_training}
\KwIn{Environment $env$; Actor $\pi_\theta$; Critic $V_\phi$; PPO clipping coefficient $\epsilon$; GAE hyperparameter $\lambda$; discount factor $\gamma$; training iterations $iters$; trajectory length $steps$; number of epochs $epochs$; Critic loss weight $vf\_coef$; entropy loss weight $ent\_coef$; batch size for parameter updates $batches$; optimizer $optim$}.
\KwOut{Policy network $\pi_\theta$}
\For{$it=1$ to $iters$}{
Interact with the environment for $steps$ time steps, collect a batch of trajectories $trajs$;\\
\For{$t=1$ to $steps$}{
$(S_t, R_t, S_{t+1})\gets trajs(t)$;\\
Calculate GAE  $\hat{A}_t$ by Eq.  \eqref{advantage};
}
\For{$i=1$ to $epochs$}{
            $t\gets 1$\;
            \While{$t \leq steps$}{
                \For{$j=1$ to $batches$}{
                    Calculate the policy ratio $r_t(\theta)$ by Eq. \eqref{policy_ratio}\;
                    Calculate the 
                    loss $\mathcal{L}_{\pi}(\theta, t)$ by Eq. \eqref{clipped_loss}\;
                    $t\gets t+1$\;
                }
                Calculate the policy loss term $\mathcal{L}_{\pi}(\theta)$ by Eq.\eqref{policy_loss}\;
                Calculate the 
                loss term $\mathcal{L}_{V}(\phi)$ by Eq. \eqref{critic_loss}\;
                Calculate the loss term $\mathcal{L}_{ent}(\theta)$ by Eq.\eqref{entropy_loss}\;
                Calculate the loss by Eq.\eqref{actor_critic_loss}\;
                $optim.\text{Step}(\theta, \phi, \Delta_{\theta,\phi}\mathcal{L}(\theta, \phi))$\;
                $t\gets t + batches$\;
            }
        }
    }
\end{algorithm}

The RAPPO algorithm is proposed to optimize the network parameters of both the Actor and Critic networks in Algorithm \ref{algo:ppo_training}. 
It collects a batch of trajectories by the previous policy during each iteration. 
Then, 
it calculates the advantage function values $\hat{A}_t$ by the Generalized Advantage Estimation (GAE) method \cite{SchulmanMLJA1} and trains the Actor and Critic networks by gradient descent-based optimization.
The advantage function describes  the difference between the action value function and the state value function, reflecting the relative advantage or disadvantage of taking a certain action in a particular state compared to the average level. 
For resource allocation, the GAE advantage estimation $\hat{A}_t$ is defined as follows:
\begin{equation}
\hat{A}_t = \sum_{l=0}^{\infty} (\gamma \lambda)^l \delta_{t+l} \label{advantage}
\end{equation}
where $\delta_t=R_t+\gamma V(S_{t+1})-V(S_t)$ is the temporal difference (TD) error at time step $t$, $R_t$ is the reward, $\gamma$ is the discount factor, and $V(S_t)$, $V(S_{t+1})$ are the value function estimates for states $S_t$ and $S_{t+1}$, respectively. $\lambda$ is the GAE hyperparameter to control the trade-off between bias and variance, 
with a value between 0 and 1. 
When $\lambda$ is close to 0, GAE produces lower variance but higher bias. Conversely, when $\lambda$ is close to 1, it produces higher variance but lower bias. The term $(\gamma \lambda)^l$ is a discount factor that reduces the weight of future TD errors in the weighted accumulation process. 
GAE combines the weighted sum of TD errors over multiple time steps to estimate the advantages. 

The RAPPO algorithm balances bias and variance during the estimation process, enhancing the stability and convergence speed of the policy gradient algorithm.
RAPPO optimizes the Actor network by maximizing the clipped surrogate loss function, as defined in  Eq.  \eqref{policy_loss}:
\begin{equation}
    \mathcal{L}_{\pi}(\theta) = \frac{1}{batches}\sum_{t=t_0}^{t_0+batches}\mathcal{L}_{\pi}(\theta, t) \label{policy_loss} 
    \end{equation}
    \begin{equation}
     \mathcal{L}_{\pi}(\theta, t) = \hat{{E}}_t\left[\hat{A}_t\cdot \min\{r_t(\theta),\text{clip}(r_t(\theta),1-\epsilon, 1+\epsilon)\}\right] \label{clipped_loss}
       \end{equation}
 where $\theta$ is the parameter of the policy network,  $batches$ is the number of samples in a training batch, $t_0$ is the starting time step of the batch samples, and $r_t(\theta)$ is the probability ratio between the new 
 and old policies, defined by
\begin{equation}
    r_t(\theta) = \frac{\pi_\theta(a_t|s_t)}{\pi_{\theta_{\text{old}}}(a_t|s_t)} \label{policy_ratio}
\end{equation}
 where $\pi_\theta(a_t|s_t)$ is the probability of action $a_t$ given state $s_t$ when the policy parameter is $\theta$. $\pi_{\theta_{\text{old}}}(a_t|s_t)$ is the corresponding probability under the previous policy. $\epsilon$ is a hyperparameter that controls the size of policy updates and is typically set to 0.1 or 0.2. The loss function helps prevent overly large policy updates, thereby maintaining algorithm stability. By using $\min\{r_t(\theta),\text{clip}(r_t(\theta),1-\epsilon, 1+\epsilon)\}$ in the loss function, RAPPO constrains the range of the probability ratio $r_t(\theta)$ to fluctuate within the interval $[1-\epsilon, 1+\epsilon]$. 
 To ensure training efficiency and stability while reducing training costs, the policy network is updated by  a subset of the collected trajectories.

The goal of the Critic network is to accurately estimate the state value function $V(S_t)$. Mean squared error  is used to measure the difference between the value function estimates and  actual values:
\begin{equation}
    \mathcal{L}_{V}(\phi) = \frac{1}{batches} \sum_{t=t_0}^{t_0+batches} (V_\phi(S_t) - G_t)^2 \label{critic_loss}
\end{equation}
 where $\phi$ 
 are 
 the parameters of the Critic network, $V_\phi(S_t)$ denotes the value function estimate by the Critic network for state $S_t$, and $G_t$ is the accumulated reward according to Eq.  \eqref{acc_return}. The Critic loss aims to minimize the squared difference between the predicted values and the observed rewards, improving the accuracy of the value function estimation.
To prevent the policy network from prematurely exploiting the explored state space and getting stuck in local optima, 
RAPPO 
uses an entropy loss function, 
$\mathcal{L}_{\text{ent}}(\theta)$, to balance the trade-off between exploration and exploitation as:
\begin{equation}
    \mathcal{L}_{\text{ent}}(\theta) = -\mathbb{E}_{S_t\sim\rho(\cdot), A_t\sim\pi_{\theta}(\cdot|S_t)}\left[\log \pi_{\theta}(A_t|S_t)\right] \label{entropy_loss}
\end{equation}

Although the exact state distribution $\rho(\cdot)$ is unknown, Monte Carlo sampling is employed   by generating approximately instances based on the policy $\pi_{\theta}(A_t|S_t)$. By maximizing $\mathcal{L}_{\text{ent}}$, the policy network is encouraged to have a more uniform distribution of probabilities over different actions, leading to a more comprehensive exploration of the environment and facilitating the search for a global optimal policy.
Since the Actor and Critic networks share a common State Encoder during training, the gradients are backpropagated by the same loss function. Therefore, the total loss function $\mathcal{L}(\theta, \phi)$ is defined as a weighted sum of the three loss terms:
\begin{equation}
    \mathcal{L}(\theta, \phi) = -\mathcal{L}_{\pi}(\theta) + \text{vf\_coef} \mathcal{L}_{V}(\phi) - \text{ent\_coef} \mathcal{L}_{\text{ent}}(\theta) \label{actor_critic_loss}
\end{equation}
 where \text{vf\_coef} and \text{ent\_coef} are weight coefficients.


\begin{algorithm}[bt!]
\small
\caption{DARA}
\label{algo:7}
\KwIn{Task $v^w_i$}
\KwOut{Allocated DC $\mathcal{D}_k$ and 
server $\mathcal{V}^k_{j*,l*}$}
Calculate  $\hat{T}^{EST}_{w,i}$ of the task by Eq.  \eqref{task_avg_est};\\
$k\gets argmin_{k=1,\dots,M}\left\{p_k(\hat{T}^{EST}_{w,i})\right\}$;\\
$(j^*, l^*)\gets argmax_{\substack{j=1,\dots,\mu_k \ l=1,\dots,\omega^k_j}}\{\frac{\xi^k{j,l}}{P_{k,j,l}}\}$;\\
Calculate the electricity cost $C^*_{w,i}$ by Eq.  \eqref{task_cost};\\
\For{$j=1$ to $\mu_k$}{
\For{$l=1$ to $\omega^k_j$}{
Calculate  $T^{EFT}_{w,i;k,j,l}$ of task $v^w_i$ on  $\mathcal{V}^k_{j,l}$ by Eq.  \eqref{task_vm_eft};\\
\If{$T^{EFT}_{w,i;k,j,l}\leq d_{w,i}$}{
Calculate $C_{w,i}$ on $\mathcal{V}^k_{j,l}$ by Eq.  \eqref{task_cost};\\
\If{$C_{w,i}\le C^*_{w,i}$}{
$(j^*, l^*)\gets (j, l)$;\\
$C^*_{w,i}\gets C_{w,i}$;\\
}
}
}
}
\end{algorithm}

\vspace{-3mm}
\subsection{Reserve Strategy} 
When the  confidence level of the action provided by the policy network is below the threshold, it is necessary to execute a reserve strategy. A heuristic resource allocation algorithm, Deadline Assured Resource Allocation (DARA) in Algorithm  \ref{algo:7} is proposed as the reserve strategy adopted in the CCRA algorithm.
DARA first selects a DC $\mathcal{D}_k$ with the lowest electricity price at the earliest possible start time $\hat{T}^{EST}_{w,i}$ of the task. 
Then, it searches for a 
server within that DC that can meet the task's deadline constraint while minimizing the electricity cost. If no 
server
satisfies the requirements, DARA selects the 
server with the highest performance-to-power ratio. The worst-case time complexity for Algorithm \ref{algo:7}  is $O(\max_{k=1,\dots,M}\left\{\omega^k\right\})$. Since ECMWS schedules tasks based on their topological order in the workflow, DARA ensures that all direct predecessor tasks of  task $v^w_i$ are completed when allocating resources 
with the completion time $T^F_{w,i
}$. 
Based on $T^F_{w,i
}$ and $T^{avail}_{k,j,l}$ of  
server $\mathcal{V}^k_{j,l}$, DARA 
calculates the earliest start time $T^{EST}_{w,i;k,j,l}$ and earliest finish time $T^{EFT}_{w,i;k,j,l}$ of task $v^w_i$:
\begin{equation}
T^{EFT}_{w,i;k,j,l}=T^{EST}_{w,i;k,j,l}+ T^D_{w,i} \label{task_vm_eft} 
    \end{equation}
    where $T^{EST}_{w,i;k,j,l}=\max\left\{T^{avail}_{k,j,l}, \max_{v^w_{i'}\in PR^w_i}{T^F_{w,i'}}\right\}$ 
and
$T^{avail}_{k,j,l}=\max_{\substack{w'=1,\dots,N; i'=1,\dots,n_{w'}; v^{w'}_{i'}\neq v^w_i}}{x_{w',i';k,j,l}\times T^F_{w',i'}}$.\\
$T^{avail}_{k,j,l}$ is defined as the latest completion time among all tasks  allocated to $\mathcal{V}^k_{j,l}$.
For the same example, the confidence level of the Actor-Critic  framework is bigger than the threshold for workflow 2. The resource is allocated as  $v^2_{6}:V^1_{2,1}$, $v^2_{5}:V^1_{1,2}$, $v^2_{2}:V^1_{2,2}$,
$v^2_{4}:V^1_{1,1} $, $v^2_{3}:V^1_{1,1}$, and $v^2_{1}:V^1_{2,1}$. 


 \vspace{-3mm}
\section{Experimental Analysis}\label{sec5}
 \vspace{-1mm}

\begin{figure}[!htb]
	\centering
	\includegraphics[width=0.35\textwidth]{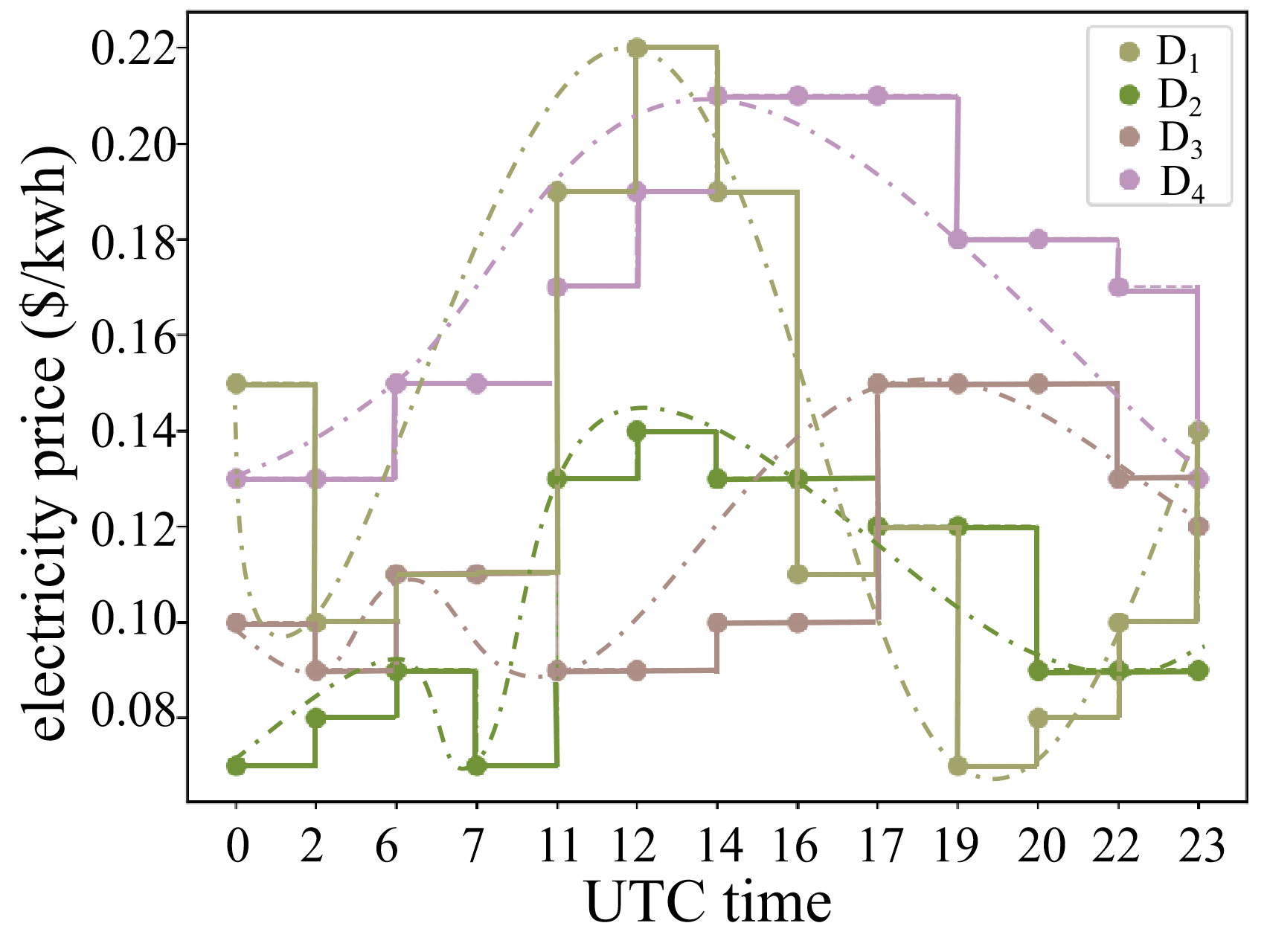}
 \vspace{-2mm}
	\caption{Electricity price fluctuations in the four DCs.}
	\label{fig:electricity_prices}
  \vspace{-3mm}
\end{figure}

The cross-domain system consists of four DCs located in  California (United States), Toronto Ontario (Canada), London  (United Kingdom), and Munich (Germany). Based on the real data in \cite{7557052}, Fig. \ref{fig:electricity_prices} shows the fluctuation of electricity prices in different DCs. 
The dashed lines depict the magnitude of the fluctuations in the electricity price of each data center. D$_1$ has the largest fluctuation and the peak and off-peak periods of electricity prices vary for each DC.

To test the adaptability of the ECMWS algorithm to the volatility of electricity prices at a larger time scale while avoiding excessive experiment costs, the termination time $T^{term}$ of the cross-domain DC system is set to one day, i.e., $3,600 \times 24 = 86,400$ seconds, and the start time is set to 0:00 UCT time. The time slot length $\tau$ is set to 10 minutes, i.e., 600 seconds.
Each DC consists of five types of clusters including CPU cores, processing speed, and power which is shown in Table \ref{tbl:host_config}. 
The number of different types of servers in each DC is randomly generated. 
Six types of 
servers with specified deployment 
cluster types are set, consistent with the assumptions of relevant studies \cite{10248241,10713971}, facilitating the determination of the power consumption of each 
server. The specific configurations are shown in Table \ref{tbl:vm_config}.

\begin{table}[!htb]
	\caption{DC 
    cluster configurations.}
   \vspace{-2mm}
	\label{tbl:host_config}
	\begin{center}
		\begin{tabular}{cccccc}
			\toprule
            Cluster Type & 
           Cluster1 & 
            Cluster2 & 
            Cluster3 & 
            Cluster4 & 
            Cluster5 \\ \midrule
			GHz/Core & 2.5 & 2.5 & 3 & 3 & 4 \\
			Cores\# & 4 & 8 & 4 & 8 & 4 \\
			MIPS/Core & 500 & 500 & 600 & 600 & 800 \\
			Total MIPS & 2,000 & 4,000 & 2,400 & 4,800 & 3,200 \\
			Power/Core & 50W & 50W & 100W & 100W & 200W \\
			Total Power & 200W & 400W & 400W & 800W & 800W \\
			\bottomrule
		\end{tabular}
	\end{center}
\end{table}

To evaluate the performance of Algorithm \ref{algo:1}, 
Relative Percentage Deviation (RPD) 
is employed to compare the electricity costs generated by different algorithm parameter and component combinations,   
calculated by:
\begin{equation}
RPD(\%) = \frac{Z - Z^*}{Z^*} \times 100\%
\label{rpd}
\end{equation}
where $Z$ represents the objective function value, i.e., the electricity cost calculated by Eq.  \eqref{obj}. $Z^*$ represents the minimum electricity cost generated by all algorithms when executing the corresponding workflow instances. 
We employ  ANOVA (Analysis of Variance) to analyze the impact of different embedding vector dimension values on the inference performance of the graph embedding models. 

\begin{table}[tb]
	\caption{
    Server configurations.}
   \vspace{-2mm}
	\label{tbl:vm_config}
	\begin{center}
		\begin{tabular}{p{25mm}p{5mm}p{5mm}p{5mm}p{5mm}p{5mm}p{5mm}}
			\toprule
            Server Type & 
            Server1 & 
            Server2 & 
            Server3 & 
            Server4 & 
            Server5 & 
            Server6 \\ \midrule
			Cores\# & 1 & 1 & 1 & 1 & 1 & 1 \\
			MIPS/Core & 500 & 250 & 600 & 300 & 800 & 400 \\
			Power/Core & 50W & 25W & 100W & 50W & 200W & 100W \\
			Deployment 
            Cluster & 1/2 & 1/2 & 3/4 & 3/4 & 5 & 5 \\
			\bottomrule
		\end{tabular}
	\end{center}
  \vspace{-5mm}
\end{table}
\vspace{-2mm}

\vspace{-3mm}
\subsection{Parameter Calibration}\label{pc}
For the considered problem in this paper,  
we follow the approach in \cite{8954766} to randomly generate  workflow applications with various sizes $N\in\left\{40,60,80,100,120\right\}$ and numbers of tasks $n\in\left\{50,100,200,500\right\}$. 
For each size and number of tasks, we generate 10 workflow instances and randomly assign arrival times $T^{submit}_w\in [0, T^{term})$. The deadlines $d_w$ for the workflow applications are determined by 
$d_w=\hat{T}^{EFT}_w\times (1+\rho) \label{exp_deadline}$.
 $\rho\in\left\{0.2,0.4,0.6,0.8,1\right\}$ is a parameter which controls the tightness of the workflow application deadlines. 
 Consequently, a total of $5\times4\times5\times10=1,000$ random workflow application instances are generated for parameter calibration.
 For Policy Network Training, we construct a reinforcement learning environment based on the cross-domain DC configuration by utilizing the OpenAI Gym framework and the extend Cloudsim Plus \cite{7987304} simulation platform. 
After experimental validations, the output dimension of the first encoder  layer is 512, and  the second encoder is 256. The first actor layer is 128 and the second actor layer is 227. The first critic layer is 128 and second critic layer is 1.  The training of the policy network parameters is performed by the RAPPO algorithm
with 100 epochs, $\gamma = 0.99$, $\lambda = 0.95$, $\epsilon = 0.1$.




\begin{table*}[t]
\footnotesize
\centering
\caption{Parameter calibaration with a 95\% Tukey HSD confidence interval.}
\label{tab:combined_task_embedding1}
\begin{tabular}{|c|c|c|c|c|c|c|c|c|c|c|c|c|c|}
\hline
\multicolumn{1}{|c|}{\multirow{2}{*}{\bf RPD(\%)}}  &\multicolumn{4}{c|}{\bf Task graph embedding vector($D_{tn}$)}&\multicolumn{4}{c|}{\bf Encoder input layer($D_{tg}$)}&\multicolumn{5}{c|}{\bf Resource graph embedding($D_{
server}$)}\\
\cline{2-14}

\multicolumn{1}{|c|}{} &  15&25&  30& 35& 35&50& 65& 70&  4& 8& 12 &16&20\\
\hline

\multicolumn{1}{|c|}{Lowerbound} &-499&-421&3003&1891 
&222&-550&1783&2699
&-130&-52&1021&1087&712
\\ \hline
\multicolumn{1}{|c|}{Uppebound} &2,098&1,723&5,031&4,098 
&2,920&1,498&3,894&4,867
&174&169&1263&1332&969\\
\hline
\multicolumn{1}{|c|}{\multirow{2}{*}{\bf RPD (\%)}}  &\multicolumn{3}{c|}{\bf CWS factor  ($\alpha_1,\alpha_2,\alpha_3$)}&\multicolumn{4}{c|}{\bf BLDP bottleneck($\beta$)}&\multicolumn{3}{c|}{\bf Task sorting ($TS$)}&\multicolumn{3}{c|}{\bf Task sorting ($Conf_{thresh}$)}\\
\cline{2-14}

\multicolumn{1}{|c|}{} &  (O.4,0.2,0.4)&(O.4,0.3,0.3)&  (O.5,0.2,0.3)& 1.5& 2&4& 8& $TS_1$&  $TS_2$& $TS_3$& 0.2 &0.5&0.8\\
\hline

\multicolumn{1}{|c|}{Lowerbound} &9.6&11.0&11.9&13.3 
&12.0&10.1&14.5&10.3
&10.4&9.4&19.0&14.6&23.0
\\ \hline
\multicolumn{1}{|c|}{Upperbound} &10.6&12.0&12.9&14.4 
&13.1&11.2&15.6&11.3
&11.4&10.4&19.9&15.5&23.9
\\ \hline
\end{tabular}
\vspace{-3mm}
\end{table*}
Based on WfCommons\cite{10.1016/j.future.2021.09.043}, 2,000 workflow instances containing 500 tasks  
are 
randomly generated for the Epigenomics, Genome, and Montage types of scientific workflows \cite{9815148}. 
 Subsequently, the dataset 
is 
randomly divided into training, validation, and testing sets in a ratio of 7:2:1. 
%
To select a suitable dimension for the task graph embedding vector  $D_{tn}$,
four  values  
$D_{tn} \in \{15, 25, 30, 35\}$, 
are
 validated, and the performance comparison results of the corresponding models are shown in Table \ref{tab:combined_task_embedding1}.
We use 
RPD to estimate the performance of different parameters, and 
a 
lower RPD represents better allocation with lower consumption of electricity cost.
We can see that $D_{tn} = 25$ 
outperforms 
the other values, 
indicating that 
it has more stable performance on  test samples. 
To determine  the output dimensions of the task graph embedding model, 
four different values, i.e., $D_{tg} \in \{35, 50, 65, 70\}$, 
are 
compared in Table \ref{tab:combined_task_embedding1}.
$D_{tg} = 50$ outperforms  better than others because of its optimal balance between complexity and expressive power. 
A 
dataset of 2,000 resource graphs 
is 
generated based on the cross-domain DC configurations by randomly setting the busy time of 
servers. 
The epoch is set as 1,000, and the batch size is set as 10.
Five different dimensions for the resource graph embedding vectors  
$D_{
server}\in\left\{4,8,12,16,20\right\}$ 
are explored, and their corresponding model performance is compared in Table \ref{tab:combined_task_embedding1}. 
We can observe that $D_{rn}=8$ exhibits significant advantages over higher-dimensional values. It demonstrates more stable performance which benefits from the proper expressive computing power. Therefore, the dimension of the resource graph embedding vector is set to 8. 

For the workflow ordering algorithm, we test  15 valid combinations of $\alpha_1,\alpha_2,\alpha_3\in\left\{0.2,0.3,0.4,0.5,0.6\right\}$ based on the weight constraint relationship  in Eq.  \eqref{cons_cws_alpha}. The parameter $\beta$ for BLDP  is selected from $\left\{1.5,2,4,8\right\}$. Three task sorting strategies, $TS_1, TS_2,$ and $TS_3$, are considered. $Conf_{thresh}$, which determines the confidence threshold for action selection, 
is 
motivated to balance the trade-off between exploration and exploitation. Given its range spans from 0 to 1, we systematically 
test 
$Conf_{thresh}$ at 0.2, 0.5, and 0.8.
For each workflow, 
$15\times 4\times 3\times 3=540$ experiments need to be conducted. The parameter calibration  requires a total of $1,000\times 540=540,000$ experiments to comprehensively compare 
different components and parameter combinations for the optimal algorithm configuration.
 The RPD values of the ECMWS algorithm are shown in Table \ref{tab:combined_task_embedding1} within a 95\% Tukey HSD confidence interval.
The performance of the CWS sorting in Algorithm \ref{algo:2} under different combinations of factor weights $(\alpha_1,\alpha_2,\alpha_3)$ is presented in  Table \ref{tab:combined_task_embedding1}. Among these combinations, $(\alpha_1,\alpha_2,\alpha_3)=(0.4,0.2,0.4)$ exhibits a significant superiority over others. 
The performance of different task sorting strategies is depicted in Table \ref{tab:combined_task_embedding1}. Among  these strategies, $TS_3$ demonstrates a significant advantage over the others because the perception of bottleneck layers with higher ranks ensures that tasks with larger ranks are executed before tasks with smaller ranks. 
The table also presents the performance of the CCRA algorithm under different confidence thresholds. Smaller confidence thresholds (e.g., 0.2) tend to prioritize the output actions of the policy network, while larger confidence thresholds (e.g., 0.8) have a higher probability of executing backup policies. It can be observed that  
RPD is the lowest when $Conf_{thresh}=0.5$  due to the  balance between the policy network and the backup policy.

\vspace{-3mm}
\subsection{Ablation Study}
\vspace{-1mm}
To better verify our contributions, we conduct ablation studies on the ECMWS method and the workload setting is the same as Subsection \ref{pc}. We primarily investigate the impacts of the Workflow Sequencing module, the Task Sequencing module and the RAPPO module. The 
findings, as illustrated in 
Table \ref{tab:ablation}, collectively demonstrate the importance of these modules in minimizing scheduling costs by
ECMWS.
We begin with the 
complete 
ECMWS method and then remove the Workflow Sequencing module (-WS) with a random sorting strategy, which result in an approximate 18\% increase in RPD, indicating that our tailored workflow sorting  based on slack time, workload, and resource contention factors significantly reduces scheduling expenses. Next, we maintain all other modules and randomize the task sequencing process (-TS), leading to a 22\% rise in RPD. This suggests that the absence of sub-deadline partitioning in task ordering can lead to resource misallocation and potential deadline violations. Finally, replacing RAPPO (-RAPPO) with a  load balancing method, which disregards regional electricity cost variances, causes a substantial 41\% increase in RPD. This underscores the cost inefficiency when ignoring different prices. 

\begin{table}[t]
\footnotesize
\centering
\caption{Ablation Study.}
\label{tab:ablation}
\begin{tabular}{|c|c|c|c|c|}
\hline
\multicolumn{1}{|c|}{RPD(\%)} & ECMWS&-WS& -TS& -RAPPO\\
\hline
\multicolumn{1}{|c|}{Lowerbound} &10.7&12.6&13.2&15.2 
\\ \hline
\multicolumn{1}{|c|}{Uppebound} &11.8&13.7&14.3&16.3 
\\
 \hline
\end{tabular}
\vspace{-4mm}
\end{table}

\vspace{-5mm}

\subsection{Algorithm Comparison}
\vspace{-1mm}
To examine the performance of 
ECMWS, we use the Epigenomics \cite{4723958} and Genome \cite{Prakash2021MultiDependencyAT} scientific workflows. 
These workflows are suitable for  highly pipelined applications with multiple pipelines
operating on independent chunks of data in parallel. Both 
have complex structures of workflows and are widely used in the existing studies \cite{9815148,8954766}.
To the best of our knowledge, there is no existing algorithm for the problem under study. We compare the proposed algorithm to  HEFT \cite{993206}, ECWSD \cite{8954766}, and DEWS \cite{HUSSAIN2022211} which are modified for better contrast. 
ECWSD is a static scheduling algorithm designed for optimizing the electricity cost of multiple workflows, while HEFT and DEWS only consider a single workflow. 
We have made certain modifications to the three algorithms, which involve scheduling a batch of workflows that arrive within each time slot at the end of the time slot. 

\begin{figure*}[!tb]
\centering
	{\includegraphics[width = 0.75\textwidth]{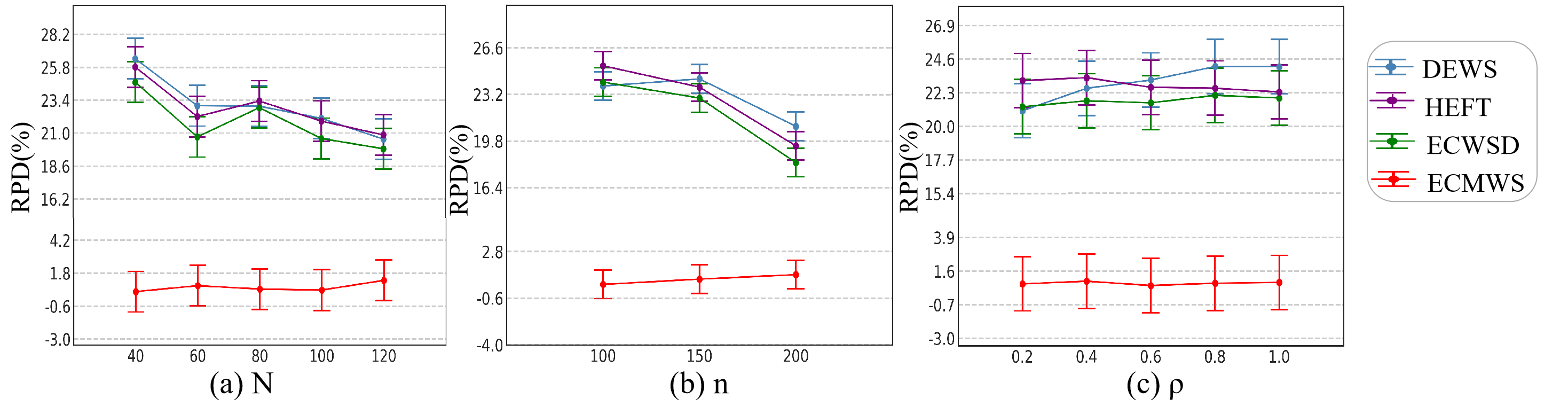}}
\caption{Algorithm comparison  for  parameters $N, n,\rho$ on  Epigenomics with 95.0\% Tukey HSD confidence level
intervals.}
\label{fig:epigenomics}
\vspace{-5mm}
\end{figure*}


Fig. \ref{fig:epigenomics}(a) presents the mean RPD  of electricity cost for Epigenomics workflows with different numbers of workflows by comparing these four algorithms. In Fig.  \ref{fig:epigenomics}(a), the ECMWS algorithm consistently outperforms the other three algorithms as the number of workflows increases. From $N=40$ to 80, the advantage of ECMWS 
over the others expands  while from $N=80$ to  120, it starts to diminish. The performance improvement ratio of the ECMWS algorithm decreases when the number of workflow instances is high. This is because that with the increasing number of workflow instances, the system undergoes more diverse state changes, which may lead to unexplored states in the strategy network. It can further lead to being trapped in local optima.  For real-world large-scale workflow deployments, to explore the performance for the proposed algorithm, we can divide 
a 
large-scale workflow into sub-workflows with the size less than 80. Fig. \ref{fig:epigenomics}(b) illustrates the mean RPD  of electricity cost for four algorithms on Epigenomics workflows with different task number configurations. DEWS, HEFT, and ECWSD are initially close but diverge as $n$ increases. 
In contrast, ECMWS demonstrates minor fluctuations and steady improvement in performance because the allocation ability of the ECMWS algorithm is improved for complex system states by graph embedding technique. 
The mean RPD of electricity cost for four algorithms  of Epigenomics workflows at different levels of workflow deadline constraints, is shown in Fig. \ref{fig:epigenomics}(c). As $\rho$ increases,  the workflow deadline constraints are relaxed. The overall performance of all algorithms improves  because  workflow tasks must be executed  on 
servers with higher processing frequencies to meet tighter workflow deadline constraints. Regardless of the degree of relaxation in the deadline constraints, 
ECMWS 
consistently outperforms the HEFT, ECWSD and DEWS algorithms. Overall, ECMWS is more reliable across different scenarios due to its consistent low-RPD performance.

\begin{figure*}[ht]
\centering
	{\includegraphics[width = 0.75\textwidth]{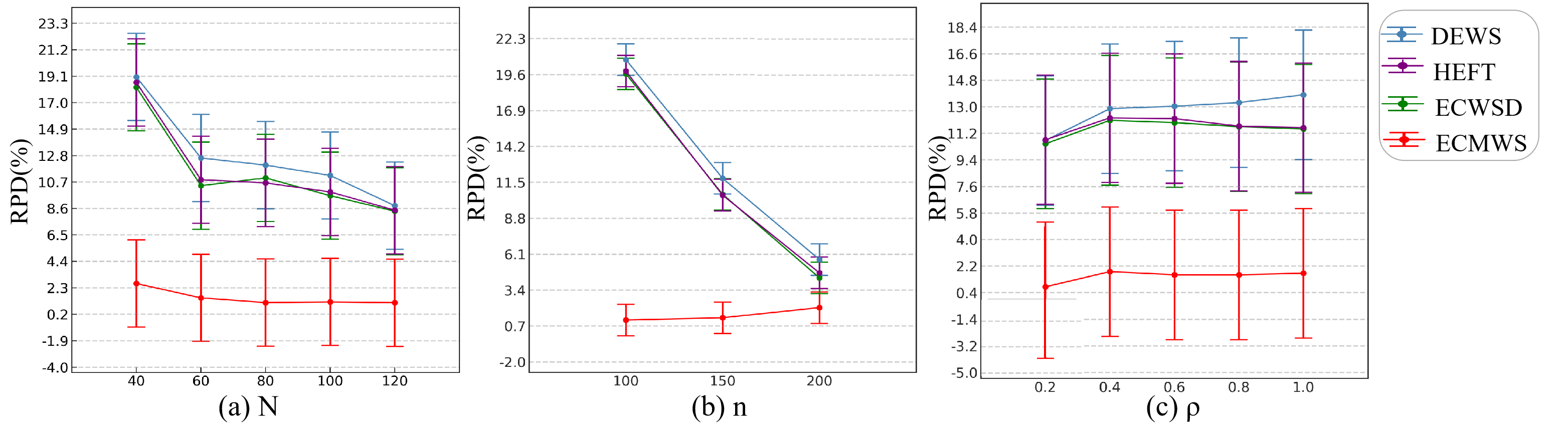}}
\caption{Algorithm comparison  for  parameters $N, n, $ on   Genome with 95.0\% Tukey HSD confidence level
intervals. }
\label{fig:Genome}
\vspace{-5mm}
\end{figure*}

For Genome workflows, Fig.\ref{fig:Genome}(a) presents the mean RPD of electricity cost with different numbers of workflows. As $N$ increases from 40 to 120, the RPD of all four methods 
declines. DEWS, HEFT, and ECMWS exhibit similar trends, with initial RPD values of approximately 19.1\%, 18.5\% and 18.0\%, respectively, when $N = 40$.
It decreases significantly when $N= 60$. Finally, it gradually declines and approaches each other. ECWSD maintains a consistently low and stable RPD throughout the change of $N$, reflecting its stability for different number of workflows. In Fig.\ref{fig:Genome} (b), as the number of tasks increases from 100 to 200, the RPD of HEFT, ECWSD and DEWS overlaps to a large extent in the early stage, while ECWSD drops more rapidly in the later stage and has the lowest RPD when $n$=200. In contrast, ECMWS exhibits minor fluctuations and consistent performance improvement as $n$ increases due
to the combination of policy networks and backup policies, which accurately predict and adjust task priorities for resource allocation. In Fig.\ref{fig:Genome} (c), as $\rho$ increases from 0.2 to 1.0, the RPD of the four methods changes relatively smoothly. The RPD  of HEFT, ECWSD and DEWS are close to each other. 
Overall, ECMWS demonstrates remarkable advantages, maintaining the lowest and most stable RPD under different parameter conditions, indicating its reliable performance with small relative deviations. In contrast, HEFT, ECWSD and DEWS show larger fluctuations in RPD values under different parameters, especially HEFT, suggesting poor stability due to its simple greedy strategy.

To compare the efficiency among these algorithms, 
Table \ref{Time} provides an overview of the electricity cost and CPU time of the comparing algorithms. According to the table, the ECMWS algorithm achieves the best performance in terms of electricity cost, albeit with a longer execution time. However, the execution time of the ECMWS algorithm falls within an acceptable range for practical implementation (seconds).

\begin{table}[t]
\centering
\caption{ Comparison on Electricity cost and CPU time.}
\label{Time}
\begin{tabular}{|c|c|c|c|c|c|}
\hline
\multicolumn{2}{|c|}{\multirow{2}{*}{\bf Algorithms}}  &\multicolumn{2}{c|}{\bf  Epigenomic}&\multicolumn{2}{c|}{\bf Genome}\\
\cline{3-6}
\multicolumn{2}{|c|}{} & \bf Cost&\bf CPU time(s) & \bf Cost& \bf CPU time(s) \\
\hline
\multicolumn{2}{|c|}{ECMWS} &0.766\%&18637.231&1.516\%&19820.938\\
\hline
\multicolumn{2}{|c|}{ECWSD} &21.755\%& 3644.691&11.531\%&4212.733\\ \hline
\multicolumn{2}{|c|}{DEWS} &23.007\%& 9416.966&12.748\%&9917.856\\
\hline
\multicolumn{2}{|c|}{HEFT} &22.824\%&188.895&11.700\% &200.195\\

\hline

\end{tabular}
\vspace{-5mm}
\end{table}

\vspace{-2mm}
\section{Conclusion }\label{sec6}
In this paper, we  address the cross-domain multi-workflow allocation problem in GDCs with fixed 
server frequency and power to minimize the electricity cost. A cost-aware multi-workflow scheduling algorithm named ECMWS  is proposed, which 
consists of the Contention-aware Workflow Sequencing, 
Bottleneck-Layer-aware sub-Deadline Partitioning,  Task Sequencing
 and  Resource Allocation. 
The experimental results demonstrate that the proposed ECMWS algorithm consistently outperforms the baseline algorithms while maintaining acceptable execution efficiency.
For future works, we will focus on exploring the cross-domain multi-workflow scheduling problem 
with multiple master nodes. For more than one masters, it needs more interactions among masters that increases the latency. The system's complexity  is increased  to manage and monitor multiple master nodes and their interactions. In addition, the topology structure of these masters would affect the algorithm complexity. 




\bibliographystyle{IEEEtran}
\bibliography{reference}
\vspace{-3mm}

\vspace{-7mm}
\begin{IEEEbiography}
[{\includegraphics[width=1in,height=1.25in,clip,keepaspectratio]{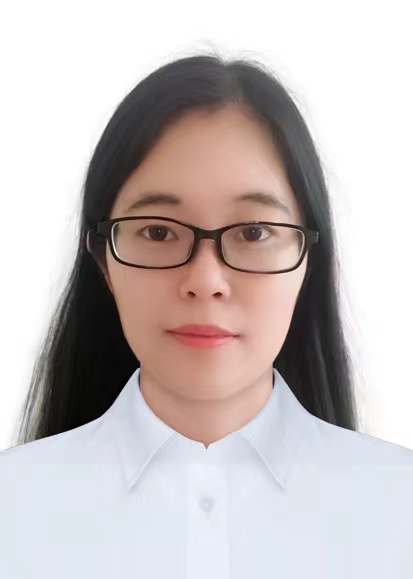}}]{Shuang Wang}
received her BSc in the College of Sciences from Nanjing Agricultural University in 2015 and completed her PhD in School of Computer
Science and Engineering, Southeast University, Nanjing, China in 2020.
She is currently a lecturer in Southeast University. Her research has been published in international journals and conferences such as \emph{IEEE Trans. on Computers, IEEE Trans. on on Parallel and Distributed Systems, IEEE Trans. on Services Computing, IEEE Trans. on Network and Service Management} and ICSOC.
Her main research interests 
include Cloud Computing, Scheduling Optimization, 
and Big Data.
\vspace{-10mm}
\end{IEEEbiography}

\begin{IEEEbiography}
[{\includegraphics[width=1in,height=1.25in,clip,keepaspectratio]{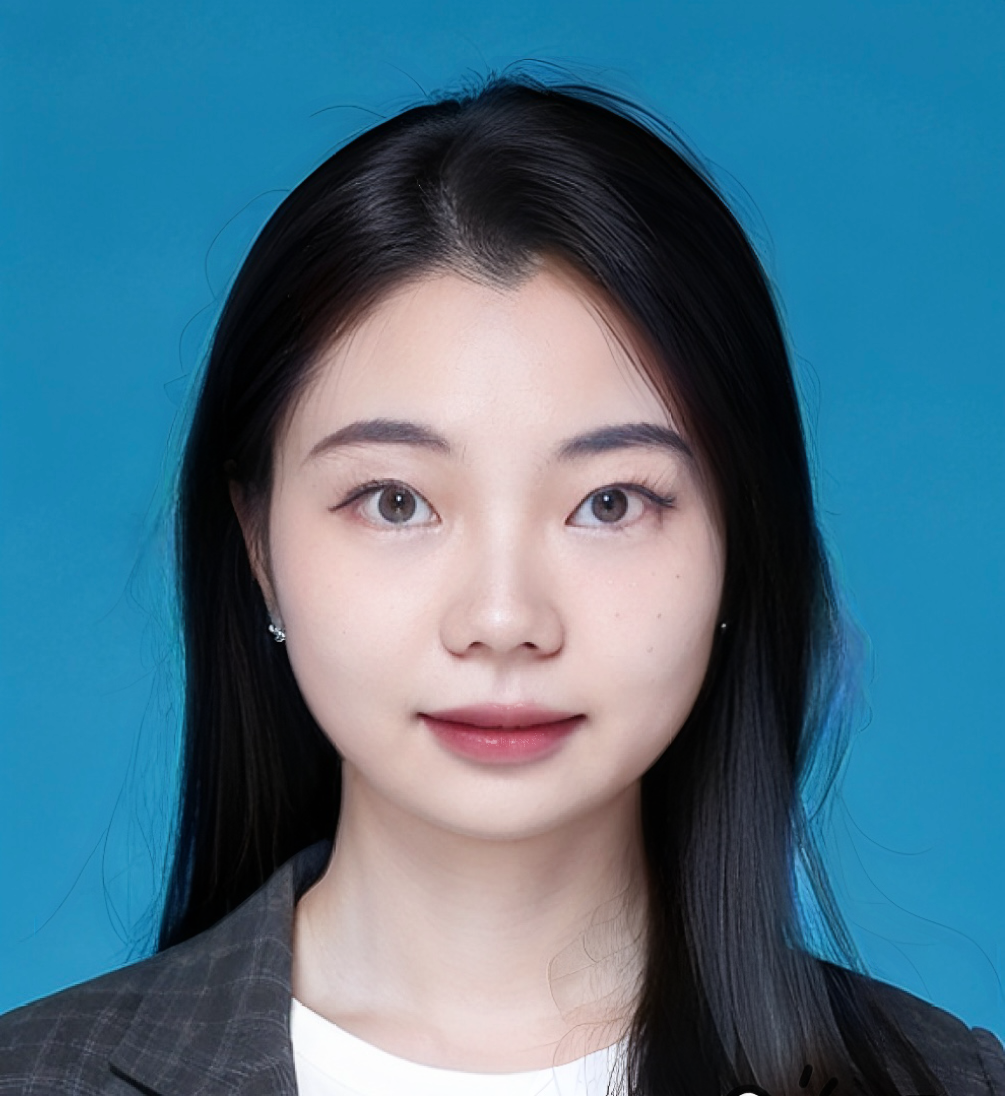}}]{He Zhang}
received her B.Sc. in Chien-Shiung WU College, Southeast University in 2022. She is currently a  Master student in School of Computer
Science and Engineering, Southeast University, Nanjing, China. Her main research interests focus on Truth Discovery and Cloud Computing. \vspace{-10mm}
\end{IEEEbiography}

\begin{IEEEbiography}
[{\includegraphics[width=1in,height=1.25in,clip,keepaspectratio]{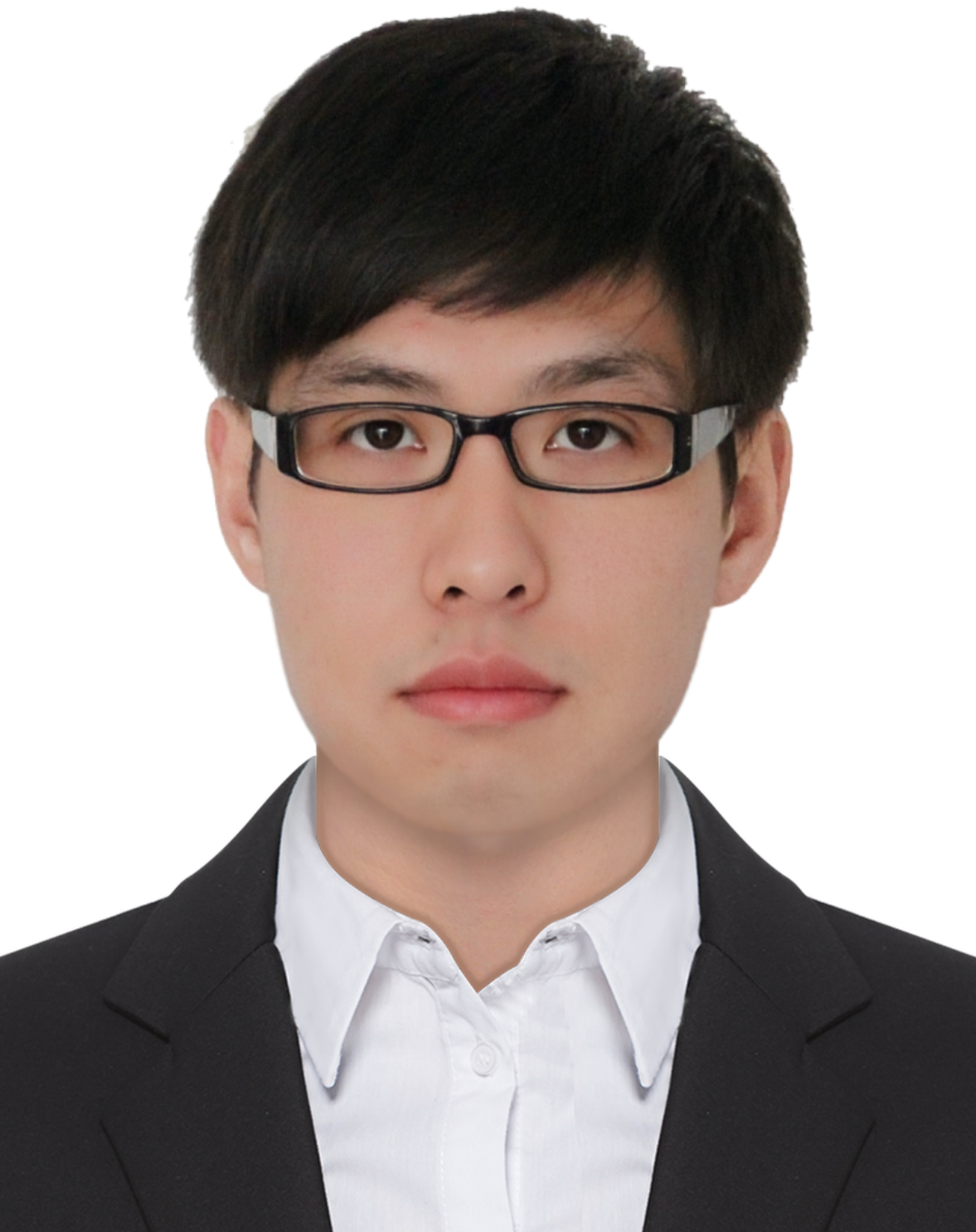}}]{Tianxing Wu}
is an associate professor working at School of Computer Science and Engineering of Southeast University, P.R. China. He received the Ph.D. degree (2018) in Software Engineering from Southeast University. He has published over 60 papers in top-tier international conferences and journals, including SIGMOD, ICDE, AAAI, IJCAI, SIGIR, TKDE and etc. His research interests include: Knowledge Graph, Knowledge Representation and Reasoning, and Artificial Intelligence Applications.\vspace{-10mm}
\end{IEEEbiography}

\begin{IEEEbiography}
[{\includegraphics[width=1in,height=1.25in,clip,keepaspectratio]{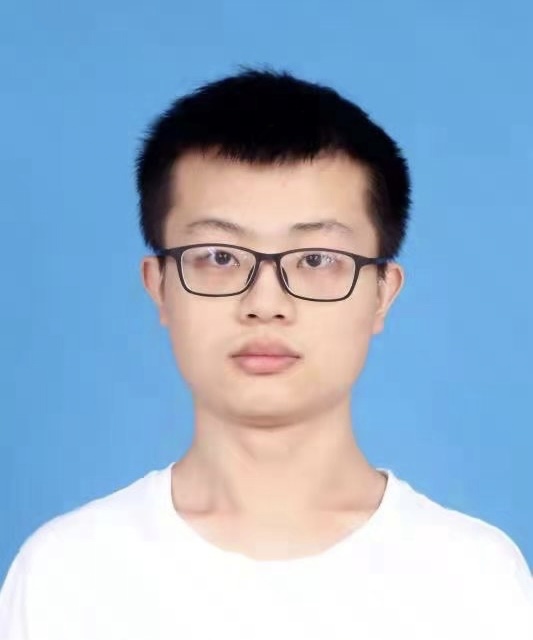}}]{Yueyou Zhang}
 received his B.Sc. in School of Computer and Communication Engineering, University of Science and Technology in Beijing in 2020. He received his master's degree in School of Computer Science and Engineering, Southeast University in 2023. His main research interest focuses on cloud computing and resource scheduling. \vspace{-10mm}
\end{IEEEbiography}


\begin{IEEEbiography}
[{\includegraphics[width=1in,height=1.25in,clip,keepaspectratio]{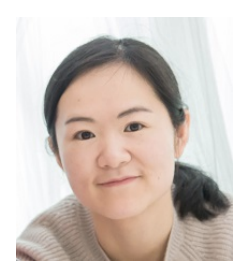}}]
{Wei Emma Zhang} is a Senior Lecturer and Associate Head of People and Culture at the School of Computer and Mathematical Sciences, 
The University of Adelaide. 
Dr Zhang is an ARC Industry Early Career Research Fellow 2023-2026.  She received her PhD from the University of Adelaide in Computer Science.  Dr Zhang's research interests include document summarization, adversarial attacks and artificial intelligence of Things. 
She has more than 100 publications 
and is among the top 100 authors worldwide, ranked by field-weighted citation impact, in the SciVal topic for Network Security. 
\vspace{-10mm}
\end{IEEEbiography}

\begin{IEEEbiography}
[{\includegraphics[width=1in,height=1.25in,clip,keepaspectratio]{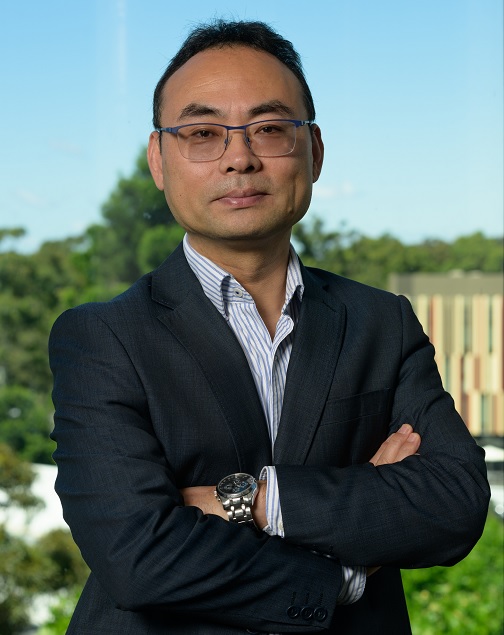}}]{Quan Z. Sheng}
is a Distinguished Professor and Head of School of Computing at Macquarie University, Australia. His research interests include Service Computing, 
Web Technologies, machine learning and Internet of Things (IoT). Michael holds a PhD degree in computer science from the University of New South Wales (UNSW).
Prof Sheng is the recipient of AMiner Most Influential Scholar Award in IoT in 2019, ARC (Australian Research Council) Future Fellowship in 2014, Chris Wallace Award for Outstanding Research Contribution in 2012, and Microsoft Research Fellowship in 2003. He is ranked by Microsoft Academic as one of the Most Impactful Authors in Services Computing (the 4th all time) 
in 2021.
Prof Sheng is the Vice Chair of the Executive Committee of the IEEE Technical Community on Services Computing 
and a member of the Australian Computer Society Technical Advisory Board on IoT.
\end{IEEEbiography}

\vfill

\end{document}